\documentclass[fleqn,usenatbib]{mnras}
\usepackage[T1]{fontenc}
\DeclareRobustCommand{\VAN}[3]{#2}
\let\VANthebibliography\thebibliography
\def\thebibliography{\DeclareRobustCommand{\VAN}[3]{##3}\VANthebibliography}
\usepackage{graphicx}	
\usepackage{amsmath}
\usepackage{verbatim}
\usepackage{xcolor}
\usepackage{xspace}
\usepackage{txfonts}
\usepackage{ulem,lipsum}
\usepackage{threeparttable}
\usepackage{comment}
\input{gridlinefig.sty}

\newcommand{\srcshort}{J0431--71\xspace}
\newcommand{\gaiacounterpart}{4654068416406146048\xspace}
\newcommand{\wisectp}{J043115.86-711728.1}
\newcommand{\RAerosita}{$04^{\rm h}31^{\rm m}15^{\rm s}.8$\xspace}
\newcommand{\DECerosita}{$-71^{\rm \circ}17^{\rm '}30\farcs0$\xspace}
\newcommand{\RAxmm}{$04^{\rm h}31^{\rm m}16^{\rm s}.24$\xspace}
\newcommand{\DECxmm}{$-71^{\rm \circ}17^{\rm '}28\farcs6$\xspace}
\newcommand{\RAgaia}{$4^{\rm h}31^{\rm m}15^{\rm s}.89$\xspace}
\newcommand{\DECgaia}{$-71^{\rm \circ}17^{\rm '}28\farcs13$\xspace}
\newcommand{\RAwise}{$4^{\rm h}31^{\rm m}15^{\rm s}.86$\xspace}
\newcommand{\DECwise}{$-71^{\rm \circ}17^{\rm '}28\farcs18$\xspace}
\newcommand{\eR}{\textit{eROSITA}\xspace}
\newcommand{\srg}{Spectrum Roentgen Gamma\xspace}
\newcommand{\xmm}{\textit{XMM-Newton}\xspace}
\newcommand{\swift}{\textit{Swift}\xspace}
\newcommand{\epn}{EPIC-pn\xspace}

\newcommand{\fluxcgs}{erg~~cm$^{-2}$s$^{-1}$\xspace}
\newcommand{\lumcgs}{erg~s$^{-1}$\xspace}
\newcommand{\rsun}{$R_{\odot}$\xspace}
\newcommand{\esrc}{\mbox{\texttt{eRASSU\,J043115.8-711730}}\xspace}

\title[Symbiotic SSS in the Magellanic Bridge]{\esrc\,:the first pulsating symbiotic supersoft X-ray source in the Magellanic Bridge}

\author[Saha et al.]{
Tathagata Saha,$^{1}$\thanks{E-mail: tathagata.saha@iucaa.in}
Chandreyee Maitra,$^{1,2}$
Philip Charles,$^{3,4}$
Frank Haberl,$^{2}$
David Kaltenbrunner,$^{2}$ \newauthor
Andrzej Udalski,$^{5}$
David Buckley,$^{6,7}$
Itumeleng Monageng,$^{6}$
Lee Townsend,$^{6,8}$
Manami Sasaki,$^{9}$\newauthor
Sara Saeedi,$^{9}$
Mara Salvato$^{2}$
\\
Inter-University Centre for Astronomy and Astrophysics, Pune University Campus, 400711, Pune, India $^{1}$\\
Max Planck Institute for Extraterrestrial Physics, Giessenbachstrasse, D-85748 Garching, Germany $^{2}$\\
School of Physics and Astronomy, University of Southampton, Southampton, SO17 1BJ, United Kingdom $^{3}$\\
Department of Physics, University of the Free State, 205 Nelson Mandela Drive, Bloemfontein, 9300, RSA $^{4}$\\
Astronomical Observatory of the University of Warsaw, Al. Ujazdowskie 4, 00-478 Warszawa, Poland $^{5}$\\
South African Astronomical Observatory, PO Box 9, Observatory, Cape Town 7935, South Africa $^{6}$\\
Department of Astronomy, University of Cape Town, Private Bag X3, Rondebosch 7701, South Africa $^{7}$\\
Southern African Large Telescope, PO Box 9, Observatory, Cape Town 7935, South Africa $^{8}$\\
Dr. Karl Remeis-Sternwarte, Erlangen Centre for Astroparticle Physics, Friedrich-Alexander-Universit\"at Erlangen-N\"urnberg, \\ Sternwartstrasse 7, 96049 Bamberg, Germany $^{9}$\\
}

\date{Accepted  2026 July 31. Received 2026 July 29; in original form 2026 July 9}

\pubyear{\the\year{}}

\begin{document}
\label{firstpage}
\pagerange{\pageref{firstpage}--\pageref{lastpage}}
\maketitle

\begin{abstract}
The Magellanic Bridge stellar population is a relic of the tidal interaction between the Large and Small Magellanic Clouds.
A comprehensive view of the evolution of the Bridge stellar population requires probing the compact remnants of stellar evolution, otherwise hidden at optical wavelengths. The all-sky survey conducted by the \eR instrument on-board the \srg observatory has discovered a significant population of
compact-object-powered systems in the Bridge using X-ray.
The candidate super-soft source \esrc (hereafter \srcshort) was discovered as a part of this campaign, and 
we present here a deeper study of this source using \xmm and SALT spectroscopy, long-term optical-infrared photometry using OGLE, ATLAS, ASAS-SN, WISE, and \textit{GAIA} data.
\srcshort is a highly variable super-soft X-ray source, classified as a red giant with the \textit{GAIA} color-magnitude diagram.
The source exhibits:
(a) a thermal X-ray spectrum with a temperature of $kT$$\sim$30~eV and a bright state luminosity of
3.2$\times 10^{37}$~erg~s$^{-1}$ in the 0.15-1~keV band,
(b) Balmer emission lines, [Fe~X] coronal line,
HeII emission, and the Bowen fluorescence blend,
(c) an optical and infrared periodicity of $\sim$500--560 days in phase with the X-ray--UV emission,
(d) a `redder-when-brighter' trend in the stellar emission with a $\sim$520 day period indicating a pulsating donor star.
We argue that the observed spectral and temporal properties in \srcshort are consistent with a high-accretion rate onto a white-dwarf via Roche-lobe overflow, making \srcshort the first symbiotic source discovered in the Bridge.
\end{abstract}


\section{Introduction}\label{sec:introduction}
The Magellanic Bridge (MCB; hereafter Bridge) is a stream of gas and stars connecting the Large (LMC) 
and Small (SMC) Magellanic Clouds \citep{irwin1990}.
Observations show that the Bridge hosts a diverse stellar population with a significant population of young stars, that are likely formed {\it in situ} \citep{irwin1990, skowron2014}.
There also exists an older population of stars in the LMC which resembles a sub-group possessing kinematic and chemical properties 
(low-metallicity) distinct from the rest of the stellar population \citep{olsen2011, besla2013}.
In addition, an older stellar population has also been detected specifically in the Bridge region \citep{bagheri2013, gaia2021}.
These observations together indicate tidal interaction between the LMC and SMC with a plausible mass transfer from LMC to SMC under the following proposed scenarios: 
(a) the Bridge formed before the Clouds became bound to the Milky Way \citep{besla2007}, or
(b) it formed during a recent interaction between the Clouds $\sim 250$~Myr ago \citep{diaz2012}.
These scenarios imply multiple degenerate channels for the Bridge formation and evolution, distinct kinematic properties of the stellar components, and non-homogeneous chemical properties in the stellar population of the LMC.

The Bridge can host X-ray bright compact objects, which are the end products of its old stellar population. A systematic large-scale survey of the Bridge in X-rays was unavailable until 2020. With the advent of the \eR telescope \citep{predehl2021} onboard the Spectrum Roentgen Gamma \citep[SRG,][]{sunyaev2021} spacecraft, the study of the X-ray bright compact objects in the Bridge becomes more systematic, enabling the study of a flux-limited sample of X-ray emitting objects in the MCB, \eR all-sky surveys \citep[eRASS,][]{merloni2024} covered the Bridge region and identified several potential candidates for Bridge membership. This is demonstrated by the discovery with eROSITA of the type-I X-ray burster \texttt{eRASSt J040515.6-745202} \citep{haberl2023} and the Be/X-ray binary \texttt{eRASSU J012422.9-724248} \citep{yang2026} in the Bridge, and
where the compact objects in both cases are identified as a neutron star.

This work is a part of a systematic search for compact-object powered X-ray emitting sources in the Bridge detected with \eR. We vetted the objects to confirm Bridge membership via positional cross-matching, proper motion filtering from GAIA \citep{gaia2021}, Wide-Field Infrared Survey Explorer (WISE) colors \citep{stern2012}, and photometric redshifts, matching counterparts, removal of extragalactic sources and foreground stars \citep{schneider2022}. Further details of the sample selection methodology, and overall survey results will be presented in a forthcoming work by Maitra et al. (in preparation).

Super-soft X-ray sources (SSS hereafter) are a broad class of sources powered generally by mass transfer from a donor star to a compact object most likely a white-dwarf \citep{vandenheuvel1992}, possessing diverse spectral and temporal properties \citep[][and references therein]{greiner2000, kahabka2006}.
Empirically, their spectra can broadly be explained by a blackbody with a typical temperature of $\sim$15--80\,eV.
However, some SSS can also exhibit a hard-spectral component that can be explained by thermal Bremsstrahlung \citep[e.g.][]{maitra2022}.
The most plausible model for highly luminous super-soft X-ray emission ($L_{\rm bol} \sim 10^{36}$ -- $10^{38}$~erg s$^{-1}$) involves accretion of H-rich matter from an evolved companion star, resulting in steady nuclear burning on the white-dwarf surface \citep{vandenheuvel1992, starrfield2004, kato2010}.
SSS can exhibit periodicity in their variability on a diverse range of timescales.
For close-binary SSS (CBSS) such as CAL 83, CAL 87 etc. \citep[][and references therein]{greiner2000, kahabka2006}, periods can range from hours to a few days. Such short timescales of quasi-periodic variability are also seen in SSS exhibiting nova outbursts.
For symbiotic SSS like AG Dra, SMC-3 etc. \citep[][and references therein]{orio2007}, modulations are mostly seen in the optical/IR, and typically range from a few hundred days to a few years.

In this work, we analyze the rich dataset from our multi-wavelength follow-up campaign on the SSS \esrc discovered in the \eR all-sky surveys, to establish its physical properties. We also consider its properties in the context of the diverse population of long-period X-ray-bright SSS  \citep[e.g.][]{orio2007, luna2013, kato2013}, and discuss the implications of this work in the context of stellar evolution in the Bridge.

The paper is organized as follows. In Section \ref{sec:detection}, we discuss the associated multi-wavelength counterparts of the \eR detection of \esrc. Section \ref{sec:observations} describes the entire available dataset related to \esrc that have been analyzed in this work. Sections \ref{sec:cmd}, \ref{sec:op-ir-variability}, \ref{sec:xray-spectral-analysis}, and \ref{sec:optical-spec} describes our analysis of multi-wavelength data.
We discuss the results and characterize the nature of the source in section \ref{sec:discussion} and summarize our conclusions in section \ref{sec:conclusion}.

Throughout this work the extrinsic absorption within the Galaxy in the source-direction is adopted to be $N_{\rm H, Gal} = 8.6\times10^{20}$~cm$^{-2}$ calculated from \citet{dickey1990}. The distance of the source is assumed to be 50 kpc \citep[e.g.][]{subramanian2009}. The uncertainty on any parameter derived from our X-ray spectral fitting corresponds to the 90\% confidence limit, unless stated otherwise. Upper limits from X-ray spectral fit posteriors, corresponds to the 0.95th quantile unless stated otherwise. All X-ray analysis and reductions are performed using \textsc{Heasoft} version 6.34.

\begin{figure*}
    \centering
    \includegraphics[scale=0.44]{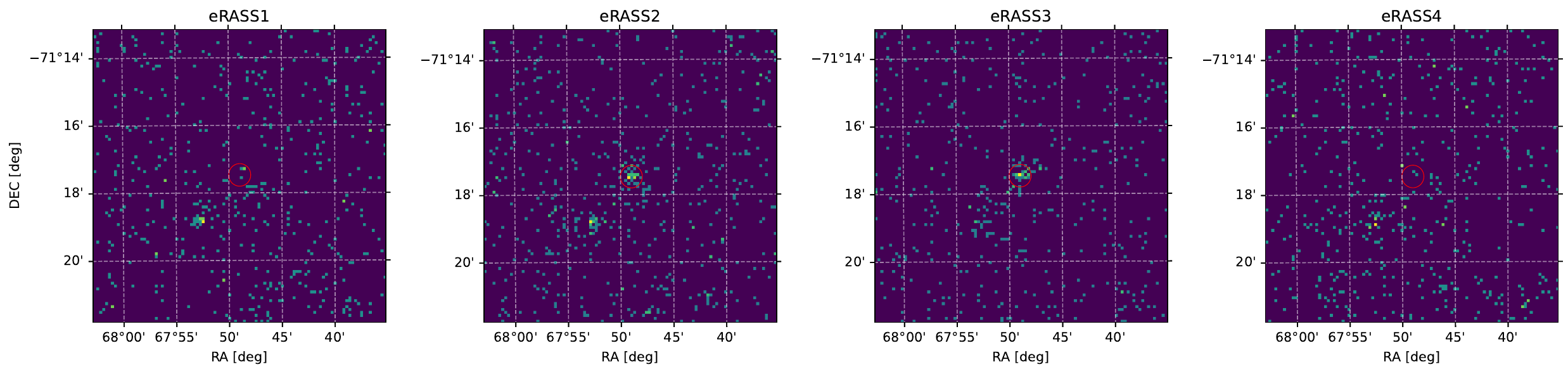}
    \caption{The four \eR all-sky scans demonstrating the X-ray variability of \srcshort over two years in the 0.2--8.0~keV energy range. The red circle (radius 30\arcsec) is centered around \srcshort.}
    \label{fig:erass-4}
\end{figure*}

\section{Detection and counterparts}\label{sec:detection}
\esrc was detected in the second and third \eR all-sky scans at $\alpha_{\rm eR}$=\RAerosita, $\delta_{\rm eR}$=\DECerosita (Figure \ref{fig:erass-4}) with a positional error of $\sigma_{\rm eR} = 1.76\arcsec$. The GAIA counterpart \gaiacounterpart\ of eDR3 \citep{gaia2021} is located at $\alpha$ = \RAgaia~, $\delta$ = \DECgaia, at an angular separation of $\sim$0.5\arcsec from the \eR position. The WISE counterpart \wisectp is located at $\alpha$=\RAwise, $\delta$=\DECwise, with a separation of $0.35\arcsec$. All optical--IR counterparts are positionally consistent within the \eR error circle. Our deeper \xmm\ follow-up observation (Section \ref{sec:xmm}) detects the source at $\alpha_{\rm XMM}$=\RAxmm, $\delta_{\rm XMM}$= \DECxmm, with a total positional error (statistical and systematic) of $\sigma_{\rm XMM}$=1.08$\arcsec$. The \xmm position is offset by $2.2\arcsec$ from the \eR position, consistent with the quadrature sum of the positional uncertainties, indicating consistent pointing to the same target.

\section{Observations}\label{sec:observations}
\esrc (\srcshort hereafter) was discovered with \eR and followed up with a dedicated \xmm\, program to identify its nature. In addition the source has extensive long-term optical/IR photometry, along with UV photometry, optical spectroscopy, and kinematic measurements such as proper motion from Gaia eDR3. We describe these datasets in the following subsections.

\subsection{\eR}\label{sec:erosita-reduction}
\eR science data products were extracted from the event files using the \eR Standard Analysis Software System (\texttt{esass} version for 030 processing   \texttt{eSASSusers\_240410\_0\_3}) task \texttt{srctools} \citep{brunner2022}. The \srcshort source region was scanned a total of 3138 times between 5th May 2020 
and 12th November 2021.
The source was detected only in the second and third all-sky scans. Through the entire period of \eR pointing, the \srcshort source region was scanned for a cumulative exposure time of 2939s, accounting for vignetting and dead-time corrections.

Source spectra were extracted from a circular region of radius 20$\arcsec$ around $\alpha=$\RAerosita, $\delta$=\DECerosita. Background spectra were extracted from 8 different circular source-free regions (each larger than the source radius) ranging from 84 to 180 $\arcsec$ in radius
to collect a sufficient number of counts in the spectral channels corresponding to the 0.2-10\,keV band for reliable background spectral modeling (Section \ref{sec:xray-spectral-analysis}). 
The sum of the area of these regions ignoring the CCD gaps and bad pixels is the background scaling factor (BACKSCAL) used for later spectral analysis.
The combined source spectra are extracted from the Telescope Modules (TM) 1 to 4 and 6. TM--5 and TM--7 suffers from light-leak \citep{predehl2021} and do not currently have a reliable calibration to correct for the problem. Spectra are obtained from all valid pixel patterns, up to PATTERN=15. All the background regions were separated by a few arc minutes from the source. The background spectra for each of the eRASS were found to exhibit total counts ranging between 166 and 190.

\subsection{\xmm}\label{sec:xmm}
\srcshort has been observed with \xmm \citep{jansen2001} on 29th August 2024: ObsID: 0941321001 (PI: C. Maitra). EPIC-pn and EPIC-MOS instruments were operated in full-frame mode (Table \ref{tab:observation-list}). The spectra are extracted with the {\tt SAS} package (version 22.1.0), Science Analysis Software \citep{gabriel2004}, and {\tt HEASOFT} (v6.34), including the corresponding calibration files. The tasks {\tt emproc} and {\tt epproc} are used for generating linearized photon event lists from the raw EPIC data.
We remove high and variable background flares by adopting a background cut of 8~cts~s$^{-1}$ arcmin$^{-2}$ (in the 10--15\,keV range) for PN and 2.5~cts~s$^{-1}$ arcmin$^{-2}$ (in the 10-12\,keV range) for MOS to generate the cleaned event file.
With this filtering we obtain a $\sim$23~ks good time interval for EPIC-pn.

We adopted a conservative circular source extraction region of 10$\arcsec$ as we attempt to minimize the contribution from photons originating from the nearby bright star HD\,270522 that is located $\sim$30$\arcsec$ from \srcshort.
The background spectra and light-curves were extracted from 80$\arcsec$ circular regions, located a few arcminutes from the source region on the same chip. We follow the recommended flag selection of the macros {\tt XMMEA\_EP} and {\tt XMMEA\_EM}.
The low energy cut-off is set to 0.2\,keV, and the high energy cut off is fixed to 10\,keV.
The final EPIC-pn source spectrum had $\sim$78 counts, and the background spectrum extracted from a few arcmin distance was found to contain 550 counts (not scaled by the BACKSCAL ratio, 0.022, of the source and the background).

\begin{table*}
\caption{Log of X-ray observations of \srcshort}\label{tab:observation-list}
\centering
    \begin{tabular}{ccccccccccc}
    \hline
    Date & Instrument      & Abbreviation & Obs.ID & Obs. & Exposure (GTI) & Opt./UV obs.  \\
    yyyy-mm-dd &&&& mode & (ks)            & Filter bands\\
    \hline
   2020-01-15 & eROSITA & eRASS1  & -- & all-sky survey  & 0.94  & --\\
   2021-01-15 & eROSITA & eRASS2  & -- & all-sky survey  & 0.83  & -- \\
   2021-01-15 & eROSITA & eRASS3  & -- & all-sky survey  & 0.81  & --\\
   2022-01-15 & eROSITA & eRASS4  & -- & all-sky survey  & 0.83   & --\\
   2023-08-05 & {\it Swift} & Sw1 & 00016228001 & PC & 4.8   & U, UVW1 \\
   2024-08-29 & \xmm  & XMM1 & 0941321001 &FF & 22.8  & UVW2$^{*}$\\
   2026-01-15 & {\it Swift} & Sw2 & 00016228002 & PC & 2.4 &  U, UVW1 \\
   \hline
   \end{tabular} 
\begin{tablenotes}
    \item Observing Mode: FF -- full frame mode (\xmm EPIC-pn), PC -- Photon counting mode (\textit{Swift}-XRT)\\
    \item *The UVW2 photometric magnitudes derived from \texttt{omichain} pipeline are not reliable as per the photometry flag.
\end{tablenotes}
\end{table*}

\subsection{\swift}\label{sec:swift}
\srcshort was observed with Swift \citep{gehrels2004} in 2023 (ObsID: 00016228001) and 2026 (ObsID: 00016228002). The XRT observations are made in PC mode and the cleaned event files were extracted using the \texttt{xrtpipeline} version 0.13.7. The source spectrum was extracted from a circular region of radius $20\arcsec$. The background spectrum was extracted from a significantly large annulus of radii $180\arcsec$ and $300\arcsec$ centered around the source, to collect enough counts to enable accurate spectral modeling of the background.

The UVOT source and background spectra were extracted from a circular region of $5\arcsec$ and an annular region of inner and outer radii $80\arcsec$ and $100\arcsec$ respectively centered around the source. We use \texttt{uvotsource} version 4.5 to perform aperture photometry on the designated source and background regions, which yielded the AB-magnitude and flux in the U and UVW1 bands.

\subsection{Optical Spectrum}\label{sec:optical-spectrum}
\srcshort was observed with the Southern African Large Telescope \citep[SALT,][]{buckley2006} on 20 November 2025 (JD\,2461000.45236111) with a 1200s exposure. The Robert Stobie Spectrograph \citep[RSS;][]{romero-colmenero2007, crause2024} was used with the PG0900 grating, covering a wavelength range of $\sim$4050--7100\,\AA. We utilised the 2" longslit, resulting in a resolving power, R$\sim$600 at 4500\,$\AA$ and R$\sim$850 at H$\alpha$. The resulting spectrum was reduced using the SALT primary data reduction and RSS science pipelines. The final extracted and wavelength calibrated spectrum was then flux calibrated using a spectro-photometric standard from the SALT archive observed with the same RSS setup as the science target.

\subsection{Photometric and kinematic datasets}\label{sec:photometry-data}
\begin{table}
    \caption{
    Photometry of \srcshort from different surveys}\label{tab:photometric-magnitudes}
    \centering
    \begin{tabular}{lccc}
    \hline
    \hline
    Catalog & Filters & Magnitude \\
    \hline
    GAIA eDR3 \citep{gaia2021}       & $G$  & $15.997 \pm 0.006$\\
    & $BP$ & $16.81 \pm 0.03$   \\
                    & $RP$ & $15.09 \pm 0.02$   \\
    \hline
     VISTA EXtension to Auxiliary Surveys            & $J$   &  13.43\\
     \citep[VEXAS,][]{khramtsov2021} & $Y$   &  14.00  \\         
                                     & $Ks$  &  12.43 \\
    \hline
    Guide Star Catalog   & $J$  & 13.65$^a$, 12.98$^b$ \\  
    \citep{lasker2008}   & $H$  & 12.81$^a$, 12.47$^b$   \\
                         & $Ks$ & 12.53$^a$, 12.15$^b$  \\
    \hline
    NOMAD \citep{zacharias2004} & $B$  & 17.91 \\ 
                                & $V$  & 16.73 \\
    \hline
    WISE  \citep{cutri2014} & $W1$  & $12.25 \pm 0.03$\\	
                            & $W2$  & $12.35 \pm 0.03$\\
    \hline
    \hline
    \end{tabular}
    \begin{tablenotes}
    \item The magnitudes from the Guide Star catalogs are at two following epochs: $^a$--1999.82, $^b$--2015.50
    \end{tablenotes}
\end{table}
\srcshort was observed with Optical Gravitational Lensing Experiment \citep[OGLE; ID: LMC 537.15.2912,][]{udalski2015} in the $I$ band over a time span of $\sim$16\,yrs, from MJD\,55264 (2010-03-09) to MJD\,61042 (2026-01-02).
Additional long-term optical photometry is obtained from public forced-photometry surveys: Asteroid Terrestrial-impact Last
Alert System
\citep[ATLAS \footnote{\href{https://atlas.fallingstar.com/}{https://atlas.fallingstar.com/}};][]{tonry2018} ($c$ and $o$ bands),
 All-Sky Automated Survey for Supernovae \citep[ASAS-SN \footnote{\href{https://www.astronomy.ohio-state.edu/asassn/}{https://www.astronomy.ohio-state.edu/asassn/}},][]{hart2023, shappee2014} ($g$ band), and 
WISE \footnote{\href{https://irsa.ipac.caltech.edu/frontpage/}{https://irsa.ipac.caltech.edu/frontpage/}} \citep{mainzer2014} (W1 and W2 bands).
We also use $G$, $BP$, and $RP$ photometric magnitudes, along with proper-motion measurements, from the GAIA eDR3 dataset \citep{gaia2021}. 
More photometric magnitudes in the $J$, $H$, $Y$, $K_s$, $B$, and $V$  bands (Table \ref{tab:photometric-magnitudes}) are obtained from the VizieR database \footnote{\href{https://vizier.cds.unistra.fr/viz-bin/VizieR}{https://vizier.cds.unistra.fr/viz-bin/VizieR}}.

\begin{figure}
    \centering
    \includegraphics[scale=0.56]{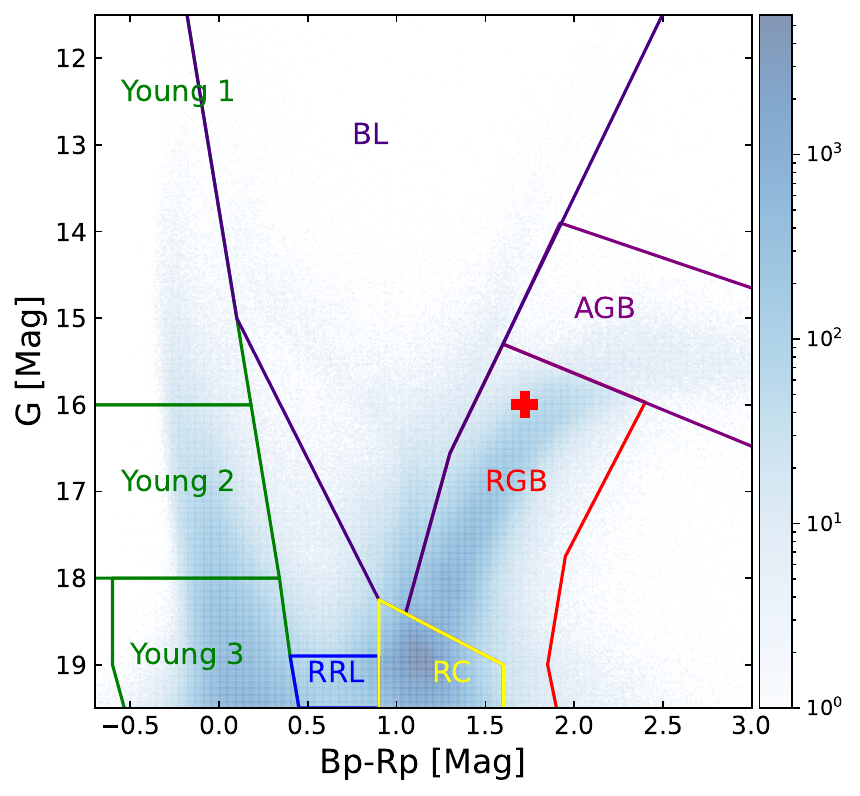}
    \caption{\srcshort Gaia CMD for stars in LMC. The bin size for the colour ($BP-RP$) and magnitude ($G$) axes are 0.01 and 0.02 respectively. The solid lines represent boundaries between different stellar types \citep{gaia2021}. J0431-71 is marked by the `+', and lies in the RGB region.}
    \label{fig:cmd-gaia}
\end{figure}

\section{colour-magnitude relations}\label{sec:cmd}
The Magellanic Clouds were covered extensively in the GAIA Early Data Release 3 \citep{gaia2021}, and we use the 
$G$, $BP$, and $RP$ photometry of all LMC stars with $G$<19.5 to construct a colour-magnitude diagram (CMD): $G$ vs $BP-RP$ (Figure \ref{fig:cmd-gaia}). The solid lines indicate boundaries between different stellar types. The shading indicates the stellar density in a given bin of the CMD.
\srcshort (Table  \ref{tab:photometric-magnitudes}) lies in the red-giant branch (RGB) region of the CMD indicating the likely presence of a late-type giant in the system. The infrared CMDs also indicate a late type red-giant (Appendix. \ref{apdx:ir-cmd}).
\\

\section{Optical and infrared variability}\label{sec:op-ir-variability}

\begin{figure*}
    \centering
    \includegraphics[scale=0.7]{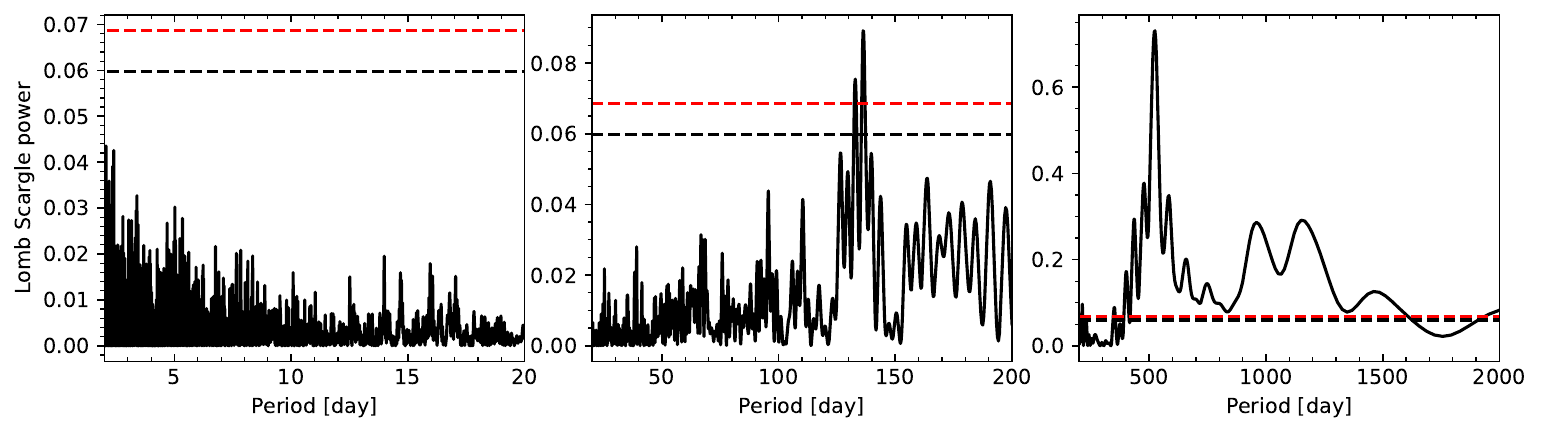}
    \caption{LS periodogram of the original OGLE I-band light-curve of \srcshort.
    The LS periodogram is plotted with three period ranges for better clarity left: 2--20 days, center: 20--200~days, and right: 200--2000~days The upper-red and bottom-black dashed lines mark the 99 per cent and 95 per cent confidence levels. The other peaks are artifacts of the sampling as verified with simulated light-curves (Section \ref{sec:op-ir-variability}).}
    \label{fig:LSperiodogramogle}
\end{figure*}
\hspace{0.05cm}
\begin{figure*}
    \centering
    \includegraphics[scale=0.78]{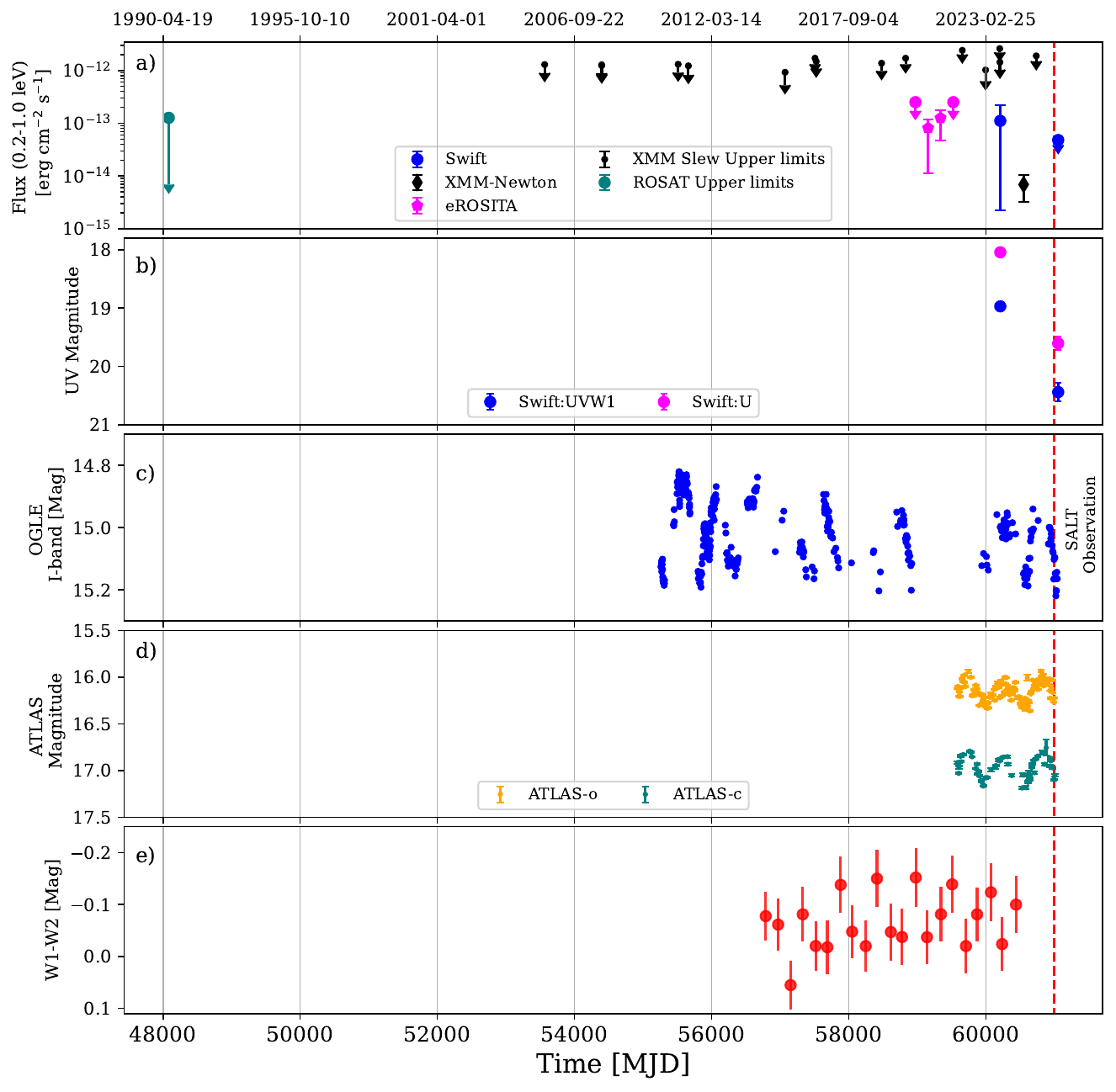}
    \caption{(a) X-ray variability including the HILIGT upper-limits (see text). These fluxes are not corrected for Galactic absorption; (b) UV photometry from \xmm and \swift; (c) long-term I-band photometry from OGLE; (d) ATLAS $c$- and $o$- band light-curves with a moving average of 10\,d; (e) variability in WISE W1-W2 colour. The red-dashed line indicates the epoch of SALT observation.}
    \label{fig:longtermlc}
\end{figure*}
\hspace{0.05cm} 
\begin{table}
\centering
    \caption{Summary of LS periodogram analysis of \srcshort}
    \begin{tabular}{lccc}
    \hline
    Facility & Band & Period ($1/f$)  & FWHM($1/f$)$\dagger$ \\
    &  & at peak power  & at peak power \\
    \hline
    &  & (day) & (day) \\
    \hline
    OGLE    & $I$   & 524.2 & 43.2\\
    OGLE    & $V$   & 506.4 & 200.0 \\
    ATLAS   & $c$ & 542.2 & 140.0\\
    ATLAS   & $o$ & 557.4 & 193.9 \\
    ASAS-SN & $g$ & 501.8 & 85.8 \\
    \hline
    \end{tabular}
    \begin{tablenotes}
   \item $\dagger$FWHM($1/f$)=$(1/f)_{\rm right} - (1/f)_{\rm left}$, where the two left and right $(1/f)$ values are frequencies corresponding to half power closest to $\nu_{\rm P-max}$ on both sides of the global maximum.
    \end{tablenotes}
    \label{tab:lsperiodogram}
\end{table}
The source exhibits variability in its optical--IR light-curve, with OGLE $I$ varying by $\sim$0.4~mag.
To search for periodicities, we compute a Lomb-Scargle periodogram \citep{lomb1976, scargle1982} \footnote{LS periodogram computation implemented through \texttt{astropy.timeseries.LombScargle} \citep{astropy2022}} for the long-term light curve. 
The confidence level corresponding to the false alarm probabilities are calculated using the method from \citep{baluev2008} implemented through the \texttt{LombScargle.false\_alarm\_level()} function of the \texttt{astropy.timeseries} module.
We obtained a maximum Lomb-Scargle power at a period of 524\,d, for the OGLE-$I$ band.
The OGLE periodogram exhibits additional peaks other than the global maximum (Figure \ref{fig:LSperiodogramogle}).
To investigate their origin, we calculate an LS-periodogram for a simulated periodic light curve over the entire period range of 2-2000 days with the same period as that of the global maximum. This light-curve is then sampled at an uniform time interval of $0.5$~days and at the same observation times as the original OGLE $I$-band data. The simulated periodogram for the uniformly sampled light-curve recovered the global periodicity peak which is used as input. However, the simulated light-curve sampled at the OGLE observation times show similar global and local maxima peaks at 213.6\,d (Figure \ref{fig:LSperiodogramogle} middle panel), 349.8\,d, 436.0\,d, 476.8\,d, 582.5\,d, 657.7\,d, 954.7\,d, and 1159.0\,d (Figure \ref{fig:LSperiodogramogle} panel on the extreme right) indicating that the local maxima are artefacts of the sampling \citep[examples by][]{vanderplas2018}.
The OGLE V-band photometry is more sparse in its sampling and gives a global period peak at 506\,d with a FWHM of 200\,d.

We obtained the ATLAS-$c$, ATLAS-$o$, and ASAS-SN $g$-band data from the forced photometry server and computed the Lomb-Scargle periodogram for each of them (Appendix \ref{apdx-lombscargleothers}). The resulting peak periods are consistent with those derived from the OGLE $I$-band light curve. However, these datasets, including OGLE-$V$ band, contain only a few cycles of the periodic variability (e.g. Figure \ref{fig:longtermlc}d for ATLAS bands), which results in a broader peak in the periodogram and hence a higher FWHM about the maximum power peak (Table \ref{tab:lsperiodogram}).
Overall the long-term light curves show maximum power at periods between 500--560\,d (Table \ref{tab:lsperiodogram}). 

WISE observes the sky in epochs separated by $\sim$six months, with each epoch consisting of multiple exposures over a short time interval. We average the light curve within each epoch to obtain one photometric point per epoch. We detect variability in the $W1-W2$ colour (Figure \ref{fig:longtermlc}), which shows a quasi-periodic behavior with successive peaks separated by $\sim$540\,d, i.e., approximately three times the WISE epoch spacing.
\section{X-ray spectral analysis}\label{sec:xray-spectral-analysis}
\begin{figure*}
    \centering
    \gridline{
    \fig{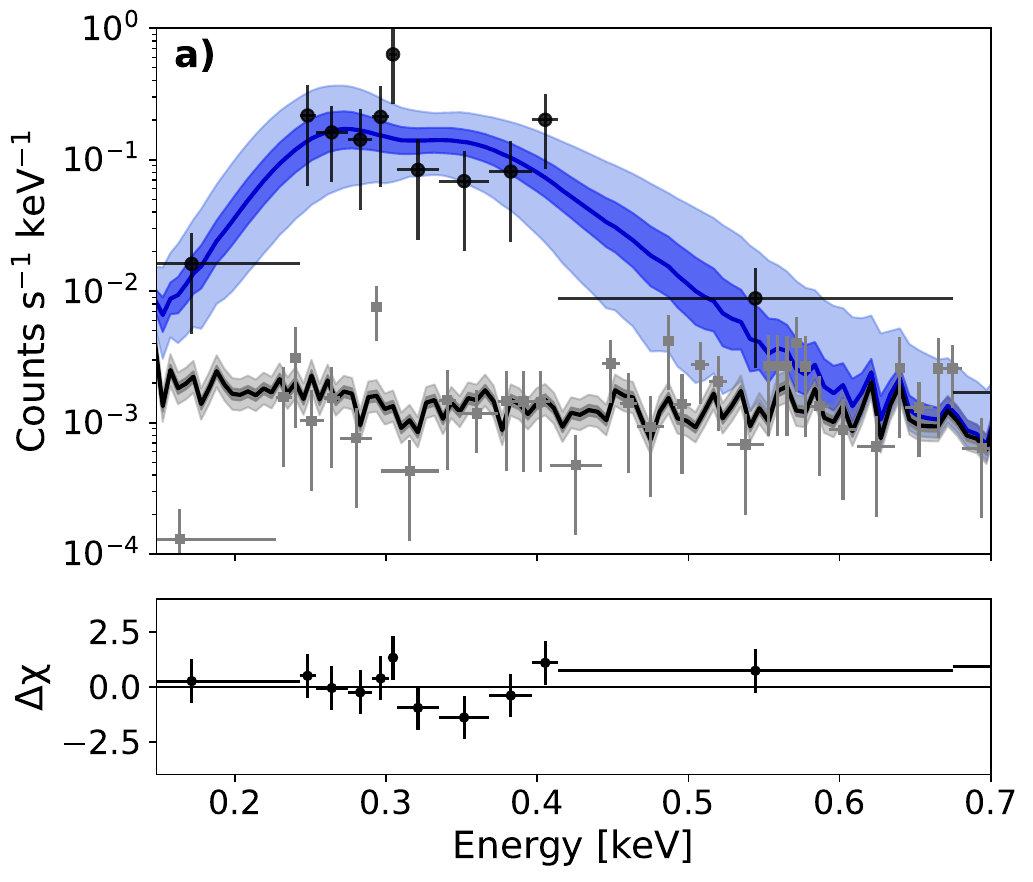}{0.335\textwidth}{}
\fig{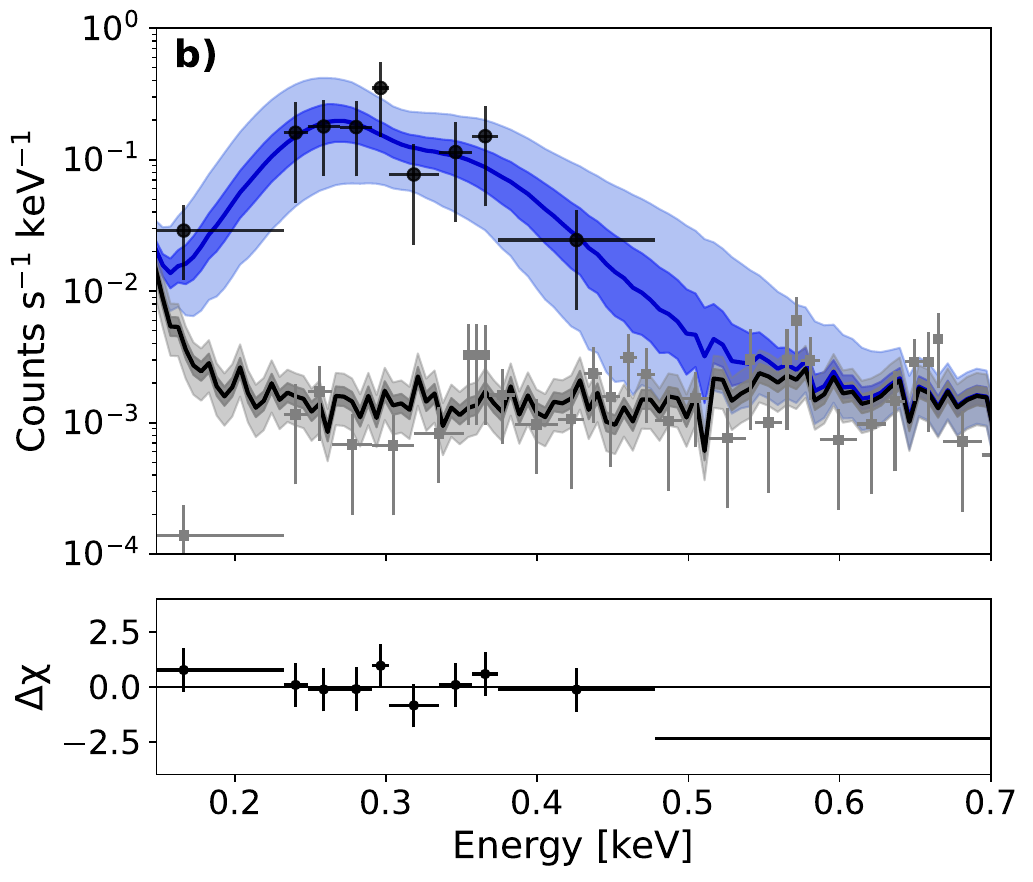}{0.335\textwidth}{}
    \fig{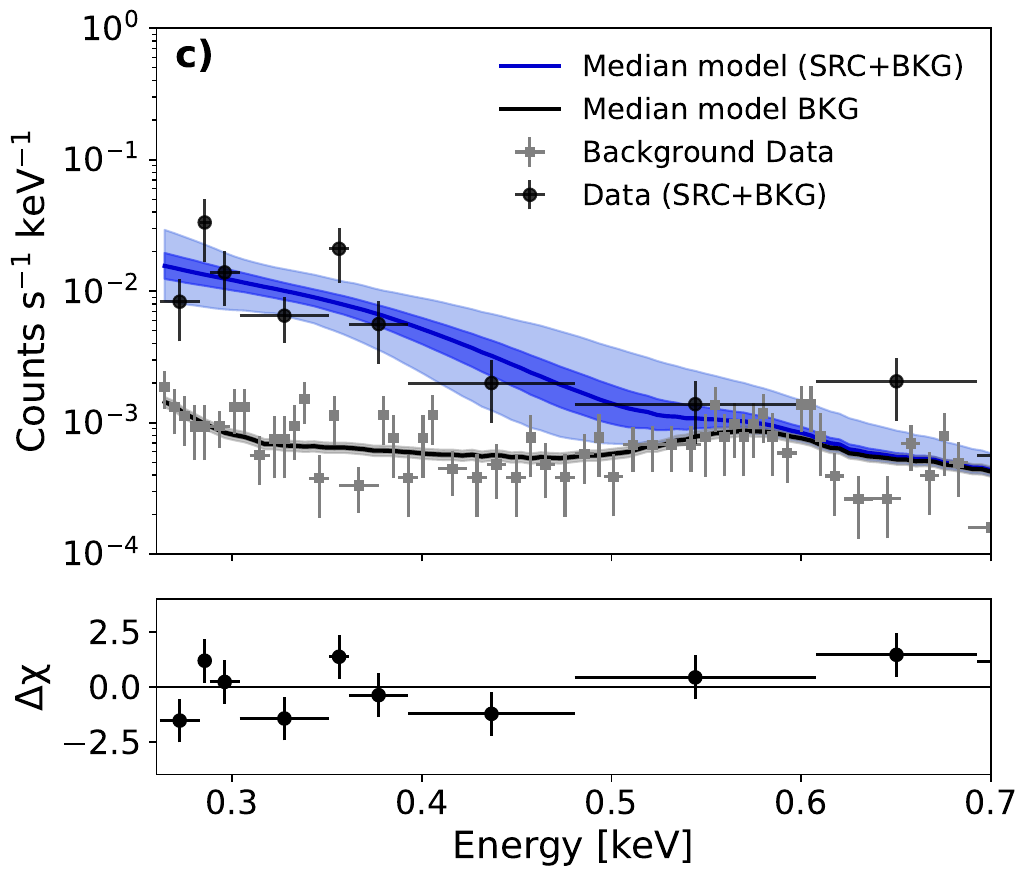}{0.335\textwidth}{} 
    }
    \caption{Convolved X-ray spectra of \srcshort from (a) eRASS2, (b) eRASS3, (c) \xmm (only \epn for clarity). The best fit source model is shown in deep blue and the best fit PCA background model is in black. The shaded regions (blue - source; grey - background) indicate 68\% (darker) and 90\% (lighter) confidence bands, with the confidence range calculated from BXA samples. \eR and \xmm spectra plotted here are grouped to 2 and 5 counts/bin respectively for visual clarity.}
    \label{fig:xray}
\end{figure*}

\srcshort is detected only in eRASS2 and eRASS3 out of the four eRASS scans.
We fit the merged spectra from the \eR telescope modules TM1-4 and 6 for each scan (Section \ref{sec:erosita-reduction}),
using Bayesian X-ray Analysis \citep[BXA,][]{buchner2014, buchner2021a}
v4.1.4 within \texttt{sherpa} \citep{siemiginowska2024} v4.18, employing \texttt{ultranest} \citep{buchner2021b}
v4.4.0 as the nested sampling algorithm. The spectra are background-dominated above $\sim$1\,keV.

To obtain reliable source parameters, we use principal
component analysis (PCA) -based background models \citep{simmonds2018} from the BXA \texttt{autobackground}\footnote{ \href{https://github.com/JohannesBuchner/BXA/tree/master/autobackgroundmodel}{https://github.com/JohannesBuchner/BXA/tree/master/autobackgroundmodel}} module in the \texttt{sherpa} environment.
The PCA-based background model is first fit to the background count spectra extracted from source-free regions.
The source spectrum is then fit simultaneously with the background model, where the background spectrum varies only in its normalization. The total model for simultaneous source and background fitting is given by:
\texttt{src\_rate\_model + $\rm \frac{bkg\_count\_model\_PCA \times BACKSCAL}{exposure\_time}$}.
Our baseline source model is \texttt{S1 = TBabs(1)*bbodyrad},
where \texttt{bbodyrad} can be treated as an approximate or phenomenological model for white-dwarf atmospheres \citep[e.g.][]{kahabka2004, orio2007} and \texttt{TBabs(1)} accounts
for Galactic absorption with $N_{\rm H}$ fixed at $N_{\rm H,Gal}$. The priors for the free parameters in the fit models are mentioned in Table \ref{tab:priors}. We also fit the data with an \texttt{apec} model, which returned similar temperature as that of the blackbody model, however was not statistically preferred based on Bayesian evidence.

We also test for the presence of an intrinsic absorption component using \texttt{S2 = TBabs(1)*TBabs(2)*bbodyrad},
where $N_{\rm H}$ in \texttt{TBabs(2)} is free.
The Bayes factor\footnote{Bayes Factor ${\rm BF_{21}} = 10^{(\ln Z_{\rm S2} - \ln Z_{\rm S1})/2.303}$},
computed relative to S1, shows that S2 is not preferred for either eRASS2 or eRASS3.
In the S2 fits, $N_{\rm H}$ is constrained to an upper limit of $\sim 5 \times 10^{20}$~cm$^{-2}$.
The blackbody temperature is $\sim$20--30\,eV for both epochs.

\begin{table}
	\centering
    \caption{Summary of priors adopted for Bayesian Spectral Analysis of the \xmm spectra}
    \begin{tabular}{lccl}
    \hline
     Parameter & Model component & Prior & Prior range \\
    \hline
     $N_{\rm H}$$^{a}$  & \texttt{TBabs(2)} & uniform & 0--1.0$^{b}$ \\
     $kT$$^{c}$ & \texttt{bbodyrad} & uniform & 1--100\\
     norm$^d$  & \texttt{bbodyrad} & Jeffreys & $10^{-3}$--$10^{16}$ \\
    \hline
    \end{tabular}
    \begin{tablenotes}
    \item (a) Only valid for \texttt{S2} see Section \ref{sec:xray-spectral-analysis}\\
    \item (b) In units of $10^{22}$~cm$^{-2}$.\\
    \item (c) In units of eV.\\
    \item (d) $R_{\rm km}^2/D_{10}^2$, ($R_{\rm km}$ = blackbody radius in km; $D_{10}$ = distance in units of 10\,kpc).
    \end{tablenotes}
    \label{tab:priors}
\end{table}
\begin{table}
    \centering
    \caption{X-ray spectral constraints on \srcshort 
    }\label{tab:xray-params}
    \begin{tabular}{ccccc}
    \hline
    Time & Obs. & $kT$ & $R_{\rm bbody}$    & Flux \\
    \hline
    58972.8 & eRASS1$^{u}$ & 30$^f$ & --  & $250.0$\\
    59158.6 & eRASS2 & $26^{+12}_{-8}$ & $0.025^{+0.32}_{-0.022}$ & $81.3_{-70.1}^{+37.6}$\\
    59339.4 & eRASS3 & $18^{+9}_{-6}$ & $0.39^{+10.49}_{-0.37} $ & $125.9^{+78.3}_{-48.3}$\\
    59525.2 & eRASS4$^{u}$ & 30.0$^f$ & --  & $260.0$ \\
    60211.0 & Sw1    & 30.0$^f$ &  0.015$\pm$0.005   & 112.7$\pm$97.6\\
    60551.5 & XMM1   & $29^{+8}_{-6}$ & $0.004^{+0.015}_{-0.003}$ & $6.67^{+3.15}_{-2.26}$\\
    \hline
    \end{tabular}
    \begin{tablenotes}
    \item mid-epoch time in MJD , kT in eV, $R_{\rm bbody}$ in $R_{\rm \odot}$, absorbed flux (0.2--2.3\,keV) in $10^{-15}$ ergs~cm$^{-2}$~s$^{-1}$.\\
	\item $^u$ upper limit on flux calculated from \eR forced aperture photometry tool.\\
    \item $^f$ frozen parameter
	\end{tablenotes}
    \end{table}
\hspace{0.05cm} 
For eRASS1 and eRASS4, we estimate flux upper limits using the \texttt{S1} model with
$kT = 30$\,eV. The total counts ($N$), background counts ($B$), and exposure ($t_{\rm EXP}$)
are obtained using the \eR aperture photometry tool \texttt{apetool} \citep{ruiz2022, georgakakis2008}.
The source count upper limit is given by
\texttt{flux\_UL = (N - B)/(ECF $\times$ EEF $\times$ t$_{\rm EXP}$)},
where the encircled energy fraction (EEF) is 0.75 and
the energy conversion factor (ECF) is $1.30 \times 10^{11}$~cm$^{2}$~erg$^{-1}$ for the \texttt{S1} model.
The resulting flux upper limits are $2.5 \times 10^{-13}$ and $2.6 \times 10^{-13}$~\fluxcgs\ for
eRASS1 and eRASS4, respectively.
For other instruments, we determined the counts from the HILIGT \footnote{\href{https://xmmuls.esac.esa.int/hiligt/}{https://xmmuls.esac.esa.int/hiligt/}}\citep{konig2022} upper limit server and used the corresponding instrument responses to calculate the ECF.

The XMM1 spectrum from the EPIC-pn, MOS1, and MOS2 instruments is fitted simultaneously using the same methodology.
We obtain a blackbody temperature of $\sim$20--30\,eV (Table \ref{tab:xray-params}) for the \xmm spectrum taken during the faint-state of the system.
Similar to the \eR spectra, the Bayes factor does not support any intrinsic absorption, with $N_{\rm H}$
from \texttt{TBabs(2)} constrained to only 
$\leq 10^{20}$~cm$^{-2}$.

The residuals show excess counts near $\sim$0.35\,keV (Figure \ref{fig:xray}).
To test its origin, we add a Gaussian component to the best-fit M1 model,
obtaining a centroid energy of $0.36 \pm 0.03$\,keV. However, an F-test of this component yields a p-value of 0.48,
indicating that the excess is consistent with Poisson noise.
We therefore adopt \texttt{S1} as the baseline model for all further calculations. 

\srcshort was detected by \textit{Swift} only in 2023 (Sw1), and we fit the resulting spectrum using the same methodology, but with $kT$ fixed at 30\,eV. We list the best-fit model parameters in Table \ref{tab:xray-params}, present the folded \eR and \xmm spectrum in Figure \ref{fig:xray} and the corresponding corner plots in Appendix \ref{apdx-bxacorners}.

\section{Optical spectroscopic analysis}\label{sec:optical-spec}
\begin{figure*}
    \centering
    \includegraphics[scale=0.72]{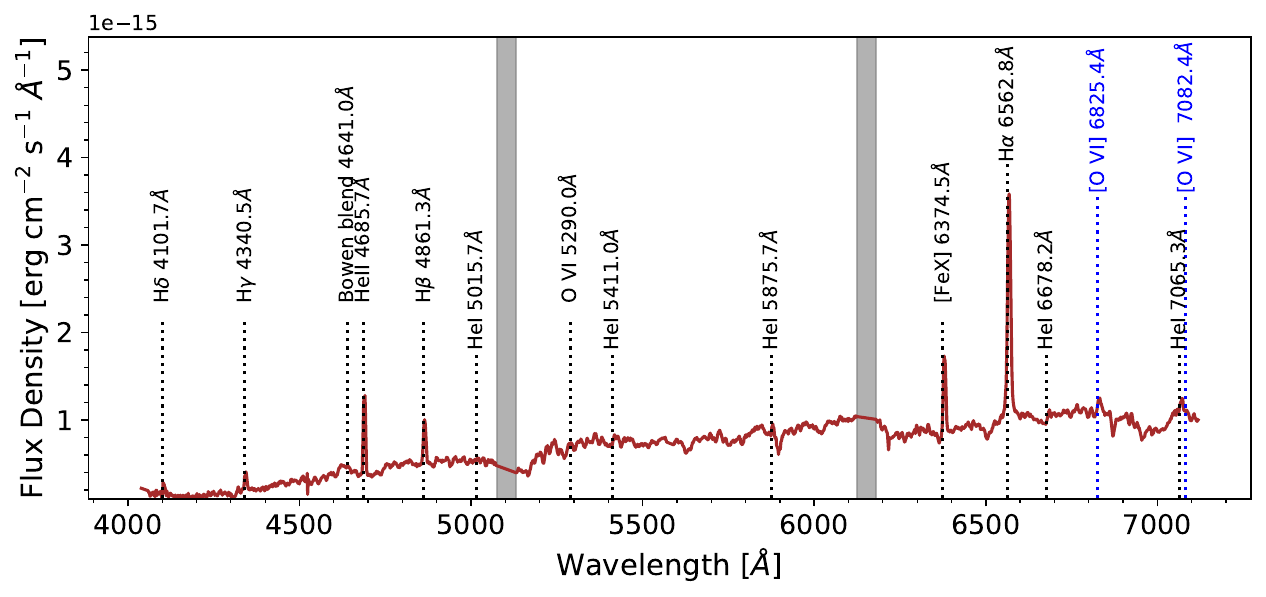}
    \caption{Flux calibrated SALT RSS spectrum from 20th November 2025. The prominent emission features have been identified. The grey lines indicate SALT chip gaps between 5075--5130\,$\AA$ and 6124--6180\,$\AA$. The weak [O VI] Raman-scattered line wavelengths are demarcated with the blue dotted lines.}
    \label{fig:optical_lines}
\end{figure*}

The optical spectrum of \srcshort exhibits Balmer emission lines, the Bowen fluorescence blend, HeII $\lambda$4686, the coronal emission line [Fe X] $\lambda$6374, and two emission line features near 6825.44\,$\AA$ and 7082.40\,$\AA$ which is consistent with the Raman-scattered emission lines \citep[e.g.][]{heo2021} of O\,VI (Figure \ref{fig:optical_lines}).
After correcting for Galactic extinction using the $E(B-V)$ value corresponding to the $N_{\rm H}$ values from \citep{dickey1990}, we characterize the most prominent emission lines by performing a phenomenological fit locally in an isolated spectral window centered on each emission line. After masking the emission line within the spectral window, the corresponding continuum is fit with a polynomial of degree $\leq 3$. The best-fit continuum is then subtracted from the spectral segment to isolate the emission line, which is subsequently modeled with a Gaussian profile (Figure \ref{fig:optical_lines}), giving the 
parameters reported in Table \ref{tab:optical-spec-parameters}.

\begin{table*}
    \caption{Parameters of the most distinct emission lines observed in the optical spectrum of \srcshort}\label{tab:optical-spec-parameters}
    \centering
    \begin{tabular}{ccccccc}
    \hline
    \hline

    Emission & Ionization & Rest  & Measured  & Line centroid & FWHM & Flux\\
    Line & Potential & Wavelength & wavelength & velocity\\
    & (eV) & ($\AA$) & ($\AA$) & (km s$^{-1}$) & (km s$^{-1}$) & ($10^{-15}$ \fluxcgs) \\
    \hline
     H$\delta$ &  13.6 & 4101.74 & $4104.76 \pm 0.37$ & $220\pm30$ & $450 \pm 68$ & $1.5 \pm 0.2$\\
     H$\gamma$ &  13.6 & 4340.47 & $4343.51 \pm 0.21$ & $210 \pm 14$ & $551 \pm 34$ & $3.0 \pm 0.2$\\
     HeII      &  54.4 &  4685.71 &  $4689.34 \pm 0.07$ & $232 \pm 4$ & $455 \pm 10$ & $12.0 \pm 0.2$\\
     H$\beta$  & 13.6 &  4861.33 &  $4864.66 \pm 0.09$ & $205 \pm 5$ &
     $482 \pm 13$ & $7.1 \pm 0.2$\\
     H$\alpha$ & 13.6 &  6562.79 &  $6567.98 \pm 0.05$ & $237 \pm 2$ &
     $517 \pm 5$ & $44.5 \pm 0.4$\\
     $\rm [Fe~X]$ & 262.1 & 6374.51 & $6379.73 \pm 0.09$ & $246 \pm 4$ & $377 \pm 10$ & $11.1 \pm 0.3$\\
    \hline
    \hline
    \end{tabular}
    \begin{tablenotes}
        \item The line fluxes are calculated by integrating the best-fit Gaussian profile.
        \item  All flux values are extinction corrected.
    \end{tablenotes}
\end{table*}

The extinction-corrected continuum spectral component, resembles a typical stellar continuum, which we model, after masking the emission lines,
using the standard stellar spectral templates from the ATLAS-T library \citep[All Spectral Type LAMOST Spectra Library,][]{ji2023},
assuming that the entire optical continuum originates from the stellar companion of the compact object.
These spectral templates are parameterized by a discrete grid of three parameters i.e. effective temperature ($T_{\rm eff}$), $\log g$, and metallicity (${\rm [Fe/H]}$).
We search for the best parameter combination within the grid to identify those for which the spectral template best matches the optical continuum.
The optimum values are obtained by minimizing the sum of squared deviations between the data and a given grid template:
$\mathcal{S}(T_{{\rm eff}, j}, \log g_{j}, {\rm [Fe/H]}_{j}) = \Sigma_{\lambda} (y_{\rm data}(\lambda) - y_{\rm model}(\lambda))^2$.
We find that this least-squares sum is primarily sensitive to $T_{\rm eff}$, with the best fit value being $T_{\rm eff} = 4040$\,K.

\section{Discussion}\label{sec:discussion}

\subsection{\srcshort: an $\alpha$/S-type symbiotic binary}\label{sec:xrayIRtype}
\srcshort exhibits three observational properties that characterize it as a symbiotic SSS: 
(a) a super-soft X-ray spectrum referred to as $\alpha$-type symbiotic \citep[e.g.][]{muerset1997, luna2013}, with spectra consistent with that observed in symbiotic SSS such as SMC~3 and Lin~358 \citep{orio2007};
(b) large optical variability with a periodicity of $\sim 524$\,d \citep{garcia1986, mikolajewska2001, mikolajewska2012}; 
(c) its position on the GAIA CMD indicating that the \srcshort binary hosts a late-type red-giant companion.
Furthermore, with the extinction-corrected IR colours indicating a lack of dust absorption (H-Ks $< 0.5$ and J-H $< 0.9$, Table \ref{tab:photometric-magnitudes}), then \srcshort is an S-type symbiotic binary \citep{corradi2008}, where the companion is a normal red giant, rather than a dust-embedded Mira variable.

\subsection{The hot super-soft component in \srcshort} \label{sec:disc-x-ray-nature}
\srcshort is a variable, symbiotic SSS, with its spectrum best described by a single blackbody of temperature in the range $(2.4$--$3.5)\times 10^5$~K. The source shows no evidence for intrinsic line of sight absorption, ionized absorption or outflow signatures in excess of the Galactic absorption column $N_{\rm H}$ value (Sec \ref{sec:xray-spectral-analysis}) across any of its epochs. This suggests that the observed variability is primarily driven by non-steady nuclear burning on the white dwarf aided by a varying mass transfer from the companion star (Section \ref{sec:disc-binary-system}). 

The unabsorbed $L_X$\,(0.15--1.0\,keV) varies between $\sim (0.02$--$3.2)\times 10^{37}$\lumcgs with the corresponding mass accretion rate $\dot{M}$ varying between $(0.01 - 1.7) \times 10^{-6}$~$M_{\odot} {\rm yr}^{-1}$ \citep[][and references therein]{hachisu2001, kato2010}. Assuming a bolometric factor of 10 \citep{orio2007}, the bolometric luminosity will vary between $L_{\rm bol} \sim 10^{36}$--$10^{38}$\lumcgs, i.e a factor of $\sim 10^2$ between the brightest and faintest states, without a corresponding significant change in the blackbody temperature (Table~\ref{tab:xray-params}). 

Such characteristic temperatures and luminosities are typical of SSS and are consistent with emission from accreting white dwarfs \citep{schaeidt1993}, where the dominant soft X-rays arise from nuclear burning of accreted material, either hydrogen \citep[e.g.,][]{vandenheuvel1992} or helium \citep{greiner2023}, on the surface of the white dwarf. Solving for the white-dwarf mass $M_{\rm WD}$ using Eq. 6 from \citet{suleimanov2003} which assumes stable or recurrent thermonuclear burning on the white-dwarf surface we obtain $M_{\rm WD}>0.6M_{\rm \odot}$ corresponding to the minimum temperature measured for \srcshort in the context of the fitted blackbody spectrum (Table \ref{tab:xray-params}). The X-ray luminosity relative to Eddington (taken to be $\sim 10^{38}$\lumcgs) ranges between 10$^{-2}$--1.0, indicating that the source transitions between sub- and Eddington accretion rate levels. 

\begin{figure*}  
    \centering
    \gridline{
        \fig{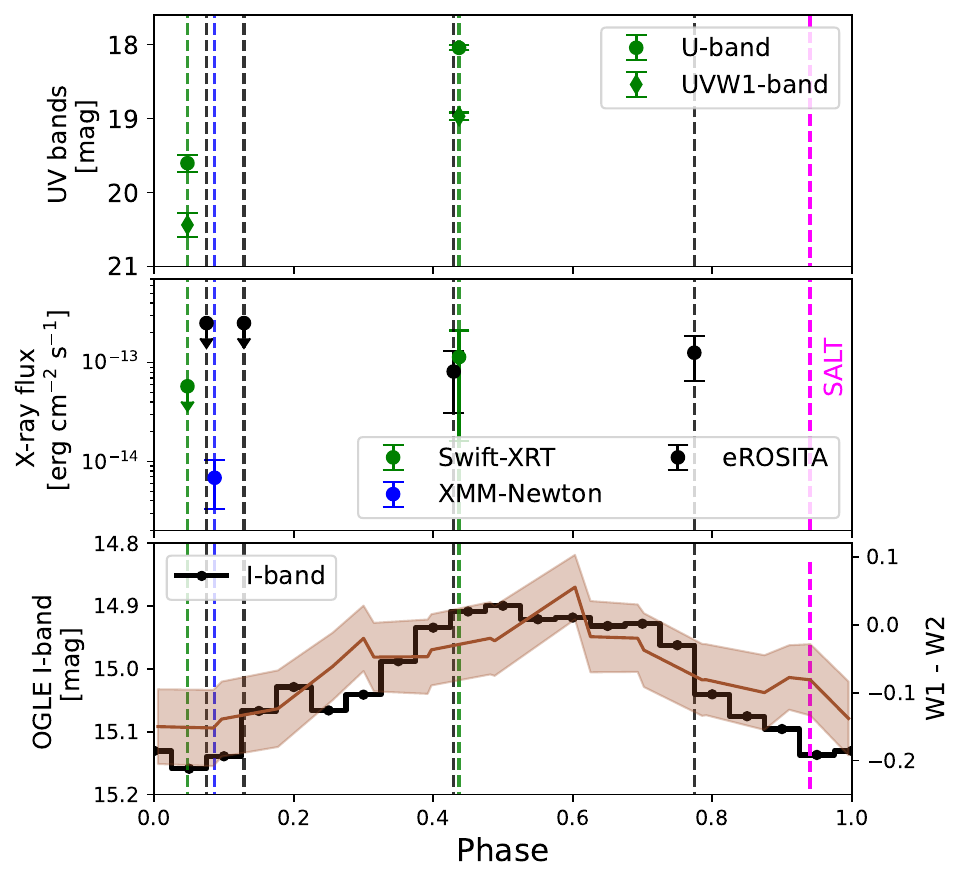}{0.52\textwidth}{(a)}
        \fig{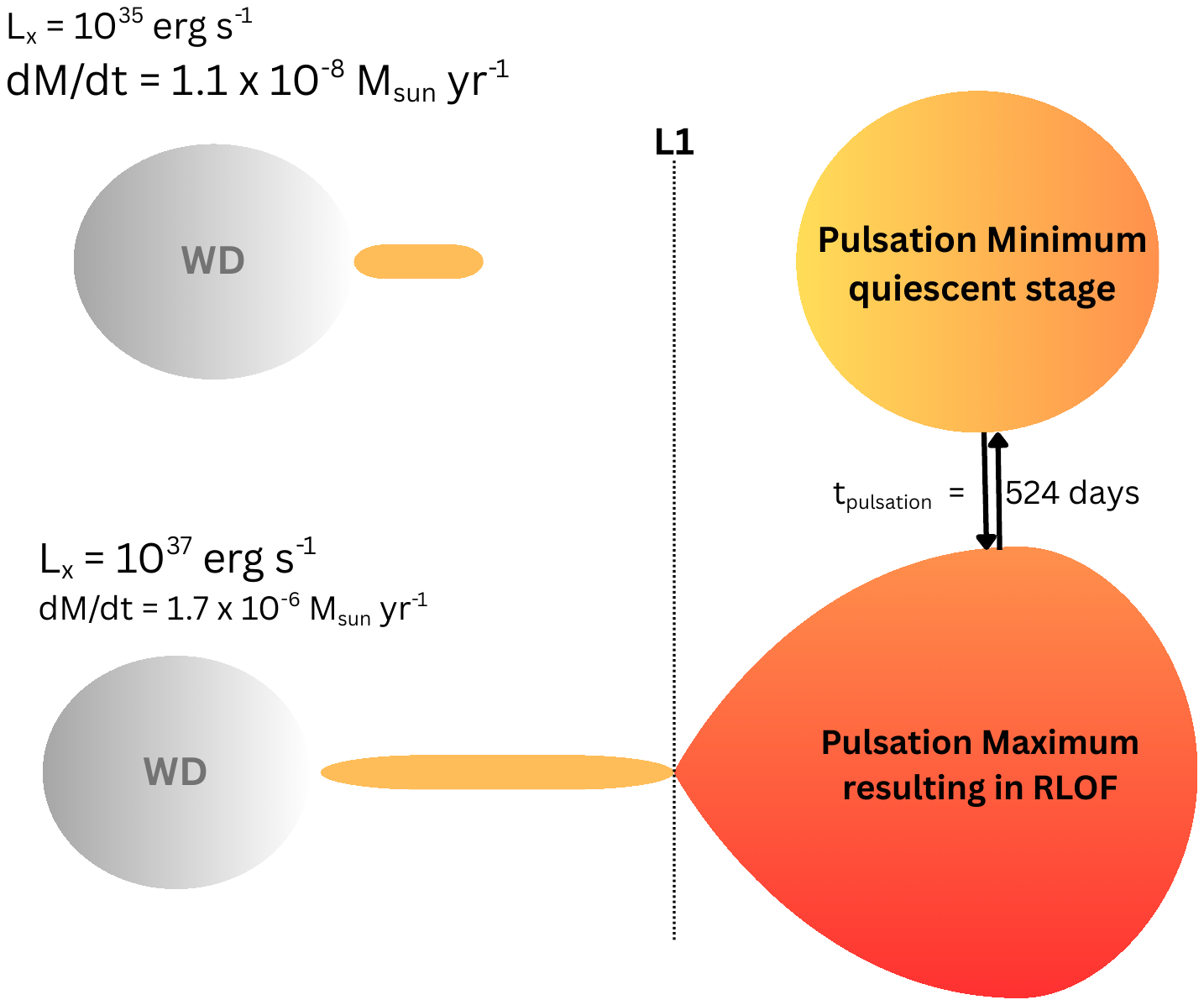}{0.52\textwidth}{(b)}}
    \caption{(a) Phased UV, X-ray, OGLE $I$ and $V$-band light-curves, and WISE colour ($W1-W2$), folded on the OGLE period of 524\,d. The brown line (and band) represents the WISE colour (and error bar). The dashed vertical lines in all panels with respective colors indicate the epochs in phase of each X-ray and the SALT observation. (b) Cartoon schematic of the effect of stellar pulsation on the accretion rate onto the white dwarf.} \label{fig:phased_multi_mechanism}
\end{figure*}

\subsection{Time variability and properties of the binary} \label{sec:disc-binary-system}
The UV and X-ray variability are in phase with the $\sim$524 day OGLE $I$-band light-curve, 
indicating that the SSS emission increases with optical brightness (Figure \ref{fig:phased_multi_mechanism}a). 
Additionally, the WISE colours show a ``redder-when-brighter'' trend.

Such long-period variability in the optical emission can, in principle, arise from either  orbital modulation of the binary system or pulsations of the donor star \citep{wood1999}. In the orbital motion driven scenario, variability could be driven by periastron-enhanced accretion via wind, which requires a non-zero eccentricity. 
However, in symbiotic systems with red-giant donors, tidal interactions can circularize the orbit, and observational studies show that systems with comparable periods are typically consistent with orbits that are near circular, or with very low eccentricity \citep{belczynski2000, fekel2000b, fekel2000a,  fekel2007}. 
More imortantly, optical and UV variability can be correlated with X-ray flux via irradiated heating of the facing side of the donor star. 
However, such processes would lead to an increase in the effective temperatures, up to 5500~K \citep[e.g.][]{kato2013} in the region irradiated by the luminous SSS component. That runs counter to our observed ``redder-when-brighter'' behaviour in the WISE colours, which instead indicate a lower effective temperature during bright phases, and which is therefore inconsistent with irradiation or reprocessing scenarios.  Instead, it is more suggestive of a stellar pulsation, where the expanded star is both brighter and cooler.

Pulsations with periods of hundreds of days are observed in red-giant systems and these are referred to as long-period variables, or LPVs \citep{wood1999,soszynski2009,soszynski2011, wood2015}. The observed periodicities in LPVs naturally match what we see in \srcshort. In this scenario, pulsation-driven changes in the stellar radius can regulate the mass transfer rate onto the white dwarf. During phases of larger radius, a Roche-lobe overflow, aided by perturbation from radiation pressure and stellar pulsation \citep[e.g.][]{dermine2009}, can lead to increased accretion with higher nuclear burning and consequently higher X-ray luminosity, thus explaining the in-phase optical and X-ray variability. Furthermore, the ``redder-when-brighter'' trend is consistent with a low-temperature but more luminous stellar emission during phases of larger effective radius of the companion star.

Overall, these observations and plausible scenarios impose the requirement of a stellar-pulsation-enhanced variable accretion onto the compact object rather than an orbital origin for the long-term variability. We can further examine the consistency of this scenario by estimating the orbital parameters under the assumption of Roche-lobe overflow from the stellar companion in a circular orbit. Using the OGLE magnitude ($I = 14.82$) and assuming a blackbody spectrum, we estimate $L_{bol} \approx 1.5\times10^4 \,L_{\odot}$ (and using a bolometric correction of $BC_{I} = -2.05$
\footnote{We use the following expression from \cite{buzzoni2010}: $BC_{I}=-2.5\log(f_{\rm bol}/f_{I}) + BC_{I, \odot}$, where,  $BC_{I, \odot}=0.64$. We assume a blackbody spectrum of temperature 3600\,K, resulting in the flux ratio $f_{\rm bol}/f_{\rm I}=11.9$. As a caveat, we caution the reader that this $BC$ estimate is strongly dependent on the spectral model assumed.}
), 
resulting in a stellar radius of $R_{\rm *} \approx 337 \,R_{\odot}$. Assuming $R_{L1} \approx R_{\rm *}$ and using the Eggleton relation \citep{eggleton1983}, we can then estimate the orbital separation and period. 

For stellar masses in the range $M_{\rm *} = 0.8$--$12\,M_{\odot}$ and $M_{\rm WD} = 0.6$--$1.2\,M_{\odot}$, the corresponding $P_{orb}$ spans $\sim400$ days (for $M_* = 12\,M_{\odot}$, $M_{\rm WD} = 0.6\,M_{\odot}$) to $\sim 2100$ days (for $M_* = 0.8\,M_{\odot}$, $M_{\rm WD} = 1.2\,M_{\odot}$). This shows that there exists combinations of masses of the binary system whose orbital periods can be comparable with the observed variability. Such a situation is possible only for systems with high total mass ($M_* + M_{\rm WD} \gtrsim 7 M_{\rm \odot}$).
We cannot directly constrain the mass of the system because the observed periodicity in the light curve is associated with stellar pulsations rather than orbital motion. 
However, several studies have estimated the masses of red-giant binaries and LPVs in the LMC (e.g. \cite{nie2017} and \cite{navarrete2025}), which
show that the mass distribution peaks around $1.5M_{\rm \odot}$, with the upper-end of the distribution extending to approximately $9M_{\odot}$. The orbital and pulsation timescales 
may become comparable for high-mass systems ($\sim 8\,M_{\rm \odot}$), however in this case, stellar-pulsation-driven accretion prevails as the dominant and self-consistent mechanism driving the observed variability (Figure~\ref{fig:phased_multi_mechanism}b), independent of the orbital parameters and, consequently, the total mass of the system. However, the total mass has implications for the relatively young age of the system and hence that it originated in the Magellanic Bridge (Section \ref{sec:bridge}).

Assuming that the accretion stream after circularization forms a slim accretion disk \citep[$H/R \leq 1$,][]{abramowicz1988} with high viscosity \citep[$\alpha\sim0.1$--0.5,][]{ss73}, given the high accretion rate during the bright phase,
it can be shown that the viscous-thermal timescale for such a disk can become comparable to the observed pulsation period (Appendix \ref{sec:apdx-accretion-timescale}), for  $M_{WD}\sim$1.0$M_{\rm \odot}$.
This estimate is a quantitative indication, that accretion streams formed during the pulsation induced mass-loss, can form and deplete within the observed pulsation cycle.

\subsection{The emission lines in \srcshort} \label{sec:disc-optical-properties}
The optical spectrum of \srcshort shows strong Balmer emission lines and HeII$\lambda$4686 with symmetric profiles, 
consistent with originating from an accretion disk. In addition, a prominent forbidden [Fe~X]$\lambda$6374 line 
($L_{\rm [FeX]} = 3.25 \pm 0.09 \times 10^{33}$~\lumcgs) is detected, 
similar to that seen in symbiotic systems located in low metallicity environments such as SMC~3 \citep{orio2007, ilkewicz2019, kato2013}. This line has an ionizing potential of 232\,eV and requires a highly ionized, optically thin medium. With the X-ray spectrum significantly strong above this energy,
the SSS becomes the primary source of the incident ionizing photons for the [Fe X] line.

The ratio $L_{\rm [FeX]}/L_{\rm X}$ varies from 0.003 in the bright 
state to 0.087 
in the faint state, indicating that the line emission remains visible strongly above the stellar continuum even as the ionizing X-ray flux declines. 
This suggests that the temporal response of the line-emitting region to the ionizing continuum is smoothed and delayed 
as a result of 
its extended spatial scale.

The critical density for [Fe~X] is $n_{\rm crit, [FeX]} = 4.8 \times 10^{9}$~cm$^{-3}$
\citep{nagao2001}. Equating this to the density of the emitter, and assuming a characteristic size $R \sim 50$--100~\rsun, the optical depth $\tau=n \sigma_{\rm T} R$ of the [FeX] emitter can be seen to range between  $\sim 0.01$--0.02, consistent with an optically thin medium. 
The above estimate of the optical depth is consistent with the requirement of an optically thin, highly ionized gas distribution for the production of the [Fe~X] emission line.
While the emission medium may be associated with such material, 
we do not observe a hard-X-ray component, although this may be due to the current spectral quality.
Otherwise, in case the hard-X-ray component is intrinsically absent, the production of [Fe X] will then be likely 
dominated by photoionizing continuum from the SSS in the energy band $E>0.26$\,keV.

We also detect weak emission features at 6830.4\,\AA\ and 7072.4\,\AA, consistent with Raman-scattered O\,VI lines.
These features are commonly observed in symbiotic systems such as AG\,Dra, CD\,43$^\circ$14304, and Z\,And \citep{schmid1999, heo2021}, and likely originate from a Raman scattering cascade of the UV O\,VI $\lambda\lambda$1032, 1038 doublet.

Together, these spectral features support the presence of an extended, photoionized, optically thin nebula 
powered by the SSS, which is consistent with symbiotic systems. Along with the observed emission lines, this indicates that 
the UV emission is likely to be strong in the spectral range not yet covered by existing data.

\begin{figure}
    \centering
    \includegraphics[scale=0.50]{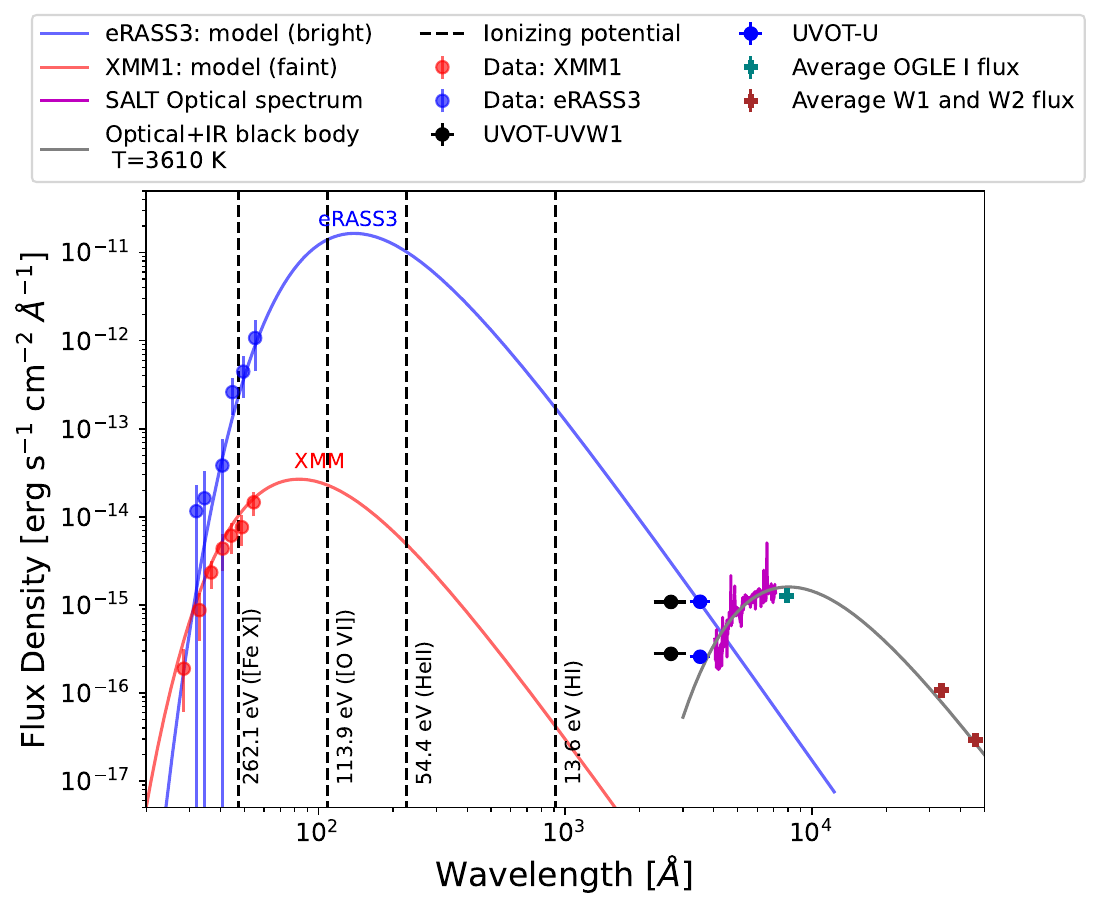}
    \caption{Broad band spectrum of \srcshort extending from X-rays to NIR, with symbols identified in the box above. All fluxes are corrected for Galactic extinction. The vertical lines indicate the ionizing potential of corresponding emission lines. For \xmm, only the \epn data is plotted.}
    \label{fig:broadband}
\end{figure}

\begin{figure*}
    \centering
    \includegraphics[scale=0.78]{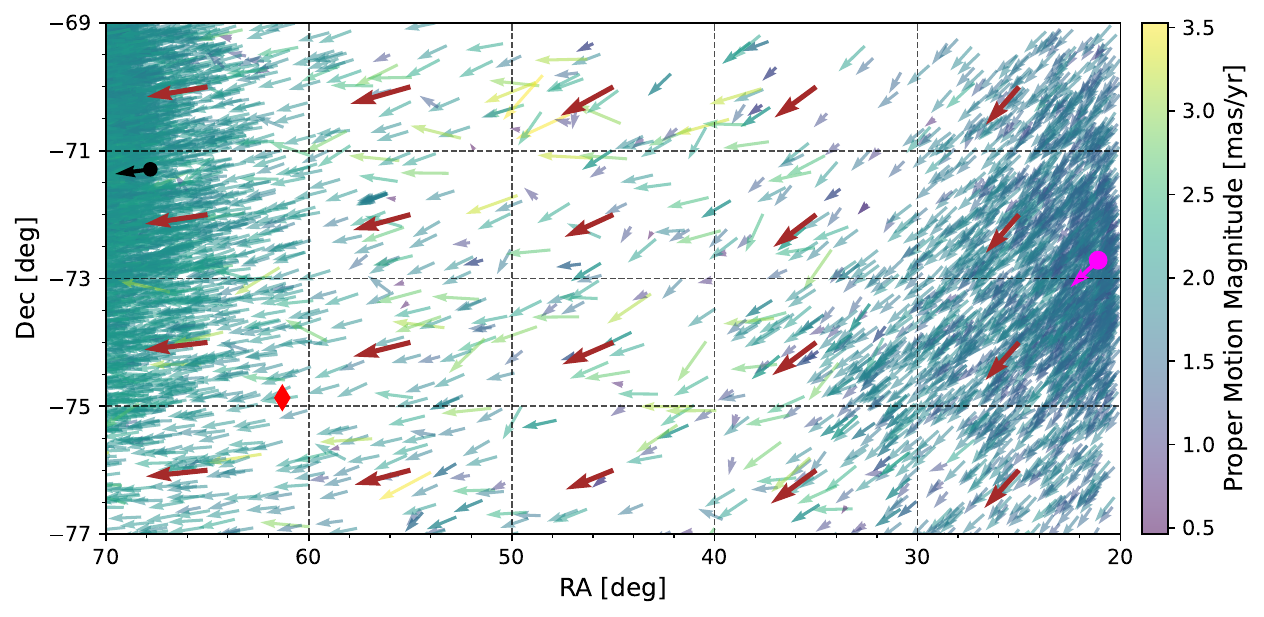}
    \caption{Proper motions of old stars in the Magellanic Bridge region from ($20^{\circ} < \alpha < 70^{\circ}$ and $-77^{\circ} < \delta < -69^{\circ}$), see \citet{bagheri2013}, cross-matched with the GAIA DR3 catalogue filtered for LMC and SMC sources. The black circle and arrow represents the position and proper-motion of \srcshort. The Magellanic Bridge burster, \texttt{eRASSt J040515.6-745202} \citep{haberl2023}, is the red diamond, and the purple circle and the arrow marks the position and proper-motion of the BeX system \texttt{eRASSU J012422.9-724248} \citep{yang2026} in the Bridge. Both the colour maps and arrow lengths for the \citet{bagheri2013} stars represent the proper motion magnitudes. The emboldened brown arrows represent the proper motion in each $10^{\circ} \times 2^{\circ}$ cell (delineated by dashed lines), and the origins are placed at each cell-centre.
    }\label{fig:bridge} 
\end{figure*}
    
\subsection{Broad band emission}\label{sec:broadband}
The broadband (X-ray--IR) spectra of symbiotic and SSS (e.g. CBSS) show significant diversity in shape. In some sources, such as RX\,J0439.8-6809, 
the extrapolated Rayleigh-Jeans tail of a single-temperature X-ray blackbody from the vicinity of the white dwarf \citep{skopal2015} 
can self-consistently explain the UV emission. 
In contrast, systems such as AG\,Dra, Lin\,358, V1974\,Cyg \citep{skopal2015}, and RX\,J0019+21 \citep{reinsch1993} show UV emission that is not reproduced by a single blackbody tail and appears relatively flat. This behaviour is commonly attributed to the presence of an additional nebular continuum component \citep{skopal2005, skopal2015}.

We construct a broadband spectral model of \srcshort using the X-ray spectra (brightest eRASS3 and faintest XMM1), UV photometry, optical spectrum, and WISE W1, W2 data (Figure \ref{fig:broadband}), with both the model and data corrected for Galactic extinction. The extrapolated X-ray blackbody towards longer wavelengths (Rayleigh-Jeans tail) does not reproduce the UV data, particularly the \textit{Swift}-U and UVW1 bands, which appear flatter than expected from a single-temperature blackbody with $kT > 15$\,eV. This suggests a different origin of the UV emission, most likely a nebular component in \srcshort similar to the other SSS symbiotics.

Since ATLAS-T templates (Section \ref{sec:optical-spec}) are not available in the IR, we model the optical spectrum and time-averaged OGLE-I, WISE W1 and W2 fluxes with a single-temperature blackbody, obtaining $T = 3600$~K. The $\sim$10\% difference from the ATLAS-T temperature is expected (Section \ref{sec:optical-spec}), as absorption in the stellar photosphere is not included in the optical+IR modeling. Overall, the broadband properties of \srcshort are consistent with other SSS, including many Symbiotic S-type SSS \citep{skopal2005}, with a temperature similar to that measured for the $\alpha$-type source Draco\,C1 \citep{aaronson1985}.

We also estimate the variation of the empirical $f_{\rm X}/f_{\rm opt}$ parameter, quantifying a relation between the emission from the star (optical) and the accretor (X-rays) in \srcshort 
and compare it with other X-ray bright symbiotic systems (Appendix~\ref{sec:apdx-betaox-methods}). The symbiotic X-ray sources collectively exhibit a harder-when-brighter trend, and \srcshort belongs to the upper-branch populated by the $\alpha$-sources, where it is consistent with this behaviour.

\subsection{J0431-71 and the Magellanic Bridge stellar population}\label{sec:bridge}
The Magellanic Bridge (MCB) was likely formed during the last tidal interaction between the LMC and SMC. 
It hosts both a young OB population \citep{irwin1990}, indicating in-situ star formation, and an older stellar population that was discovered by \cite{bagheri2013} from the 2MASS and WISE surveys.
We cross-match the old Bridge stars from the 2MASS catalogue presented in \cite{bagheri2013} with the filtered LMC and SMC catalogues from eDR3 \citep{gaia2021}, setting a maximum allowed positional uncertainty of 1\arcsec for the search. 

We obtained GAIA counterparts of $\sim 10^4$ stars from 2MASS, and plot their spatial distribution and proper motions together with the coordinates of \srcshort at $(\alpha, \delta)$=(\RAgaia, \DECgaia).
We further bin the extended Bridge region into 20 cells ($\Delta \alpha = 8.0^{\circ}$, $\Delta \delta = 2.0^{\circ}$) and compute the mean proper motion of the old stars in each cell (Figure \ref{fig:bridge}).
This depicts the evolution of the average drift-velocity in the direction from SMC to LMC.
J0431-71 lies near the periphery of the LMC (Figure \ref{fig:bridge}), close to the $\sim$90\% boundary of its stellar density distribution. Furthermore, its proper motion ($\rm \mu_{\alpha}, \mu_{\delta}) = (2.087 \pm 0.037~mas/yr, -0.275 \pm 0.038~mas/yr)$ is more aligned to the Bridge and the peripheral LMC stars, suggesting that it is more gravitationally coupled with the LMC than SMC.

Given the observed similarity of the proper motion of the old Bridge stars in the LMC periphery with the LMC stars and the absence of a mass (and consequently an age) estimate of \srcshort, we cannot conclusively determine its birth location.
The drift timescale calculated solely from the projection is $t_{\rm drift, PM} = \Delta \alpha_{bridge}/ \mu_{\rm \alpha} \sim $ (50~degree/2.09 mas/yr), which is 86\,Myr.
If \srcshort originates in the Bridge, the similarity of its proper motion to that of the LMC is explained by a sufficiently prolonged dynamical interaction with the LMC.

Therefore, the observed position and kinematics of \srcshort allow three possible interpretations of its membership: 
(a) based on its position, the source may already be part of the intrinsic old stellar population in the LMC;
(b) part of the SMC population already accreted into the LMC \citep[a distinct low-metallicity stellar population in the LMC][]{olsen2011, besla2013}
or 
(c) based on the alignment of its proper motion with that of the old Bridge stars \citep{bagheri2013, gaia2021} 
(Figure~\ref{fig:bridge}), it may also be part of the systematic drift of old stars from the SMC toward the LMC.

\section{Summary and conclusion}\label{sec:conclusion}
\srcshort was detected as an X-ray bright SSS during the eROSITA all-sky scans (eRASS:4). 
It has a GAIA optical counterpart and long-term photometric coverage from 
OGLE, ATLAS, ASAS-SN, and WISE, and we obtained optical spectroscopy with SALT. Below we 
summarize the main properties of the source and the implications of its discovery for the stellar population of the Magellanic Bridge:

\begin{itemize}

\item[a.] \srcshort is a highly variable $\alpha$-type \citep[e.g.][]{luna2013} symbiotic 
SSS with a normal red-giant stellar companion (Section \ref{sec:cmd} and \ref{sec:optical-spectrum}) as indicated by the inferred temperature of $\sim$3600~K from its optical and IR broad band spectrum and its position in the GAIA CMD.

\item[b.] The red-giant stellar companion is a pulsating LPV (Section \ref{sec:op-ir-variability}, \ref{sec:disc-binary-system}, and Table \ref{tab:lsperiodogram}) with a period of $524 \pm 43$\,d, as can be inferred from the `redder-when-brighter' trend observed in the WISE $W1-W2$ colour variation with respect to the stellar brightness (\ref{fig:phased_multi_mechanism}).

\item[c.] The soft X-ray/UV emission from the compact white-dwarf is also in phase with the stellar brightness, indicating that the mass transfer to the white-dwarf is linked to the stellar pulsation (Section \ref{sec:disc-binary-system}). These observations favour a stellar pulsation-driven accretion scenario, where a brighter, larger, and redder star undergoes a Roche-lobe overflow, driving the mass transfer rate onto the white-dwarf to $\sim10^{-6} M_{\rm \odot} {\rm yr^{-1}}$, triggering its bright X-ray state.

\item[d.] The broadband spectrum shows three distinct continuum components: the thermal X-ray component, 
the UV component, and the stellar optical--IR continuum, typical of other symbiotic SSS such as Lin\,358 and AG\,Dra \citep{skopal2015}.

\item[e.] In addition to the Balmer series, the optical spectrum exhibits Bowen fluorescence, HeII, [Fe~X], and weak Raman-scattered O~VI emission lines. Several of these lines require ionization potentials above 50\,eV, indicating the presence of a strong far-UV ionizing continuum.

\item[f.] \srcshort exhibits a strong forbidden [Fe~X]\,$\lambda$6374 coronal line ($L_{\rm [FeX]} \sim 10^{33}$~\lumcgs), similar to that in the symbiotic SMC-3 \citep{orio2007, kato2010}. The line has been observed in the X-ray faint state of \srcshort, that is, when the corresponding ionizing X-ray ($E>230$~eV) flux is significantly weak. This indicates that the line-emitting region is likely spatially extended and that its response to a highly variable ionizing continuum is rather delayed and smoothed. An extended emitter also fits well with the requirement of a tenuous optically thin medium for forbidden line production.

\item[g.] The position and proper motion of \srcshort are consistent with both the LMC and the observed systematic trend in the old stars' proper motion \citep[e.g.][]{bagheri2013, gaia2021} from SMC to LMC (Figure \ref{fig:bridge}). This observation indicates that  \srcshort may either be a part of the LMC stellar population, part of the SMC stellar population (\srcshort exhibiting SMC-3 type properties) accreted into the LMC, or part of the systematic drift of the old stellar component through the Bridge towards the LMC.

\end{itemize}

The eROSITA all-sky surveys have identified a significant population of X-ray bright sources in the Magellanic Bridge, enabling systematic studies of the stellar population and compact-object remnants in the region through multi-wavelength follow-up observations. The discovery and characterization of the symbiotic SSS \srcshort are a part of this broader time-domain effort to study the interactions between the LMC and SMC in particular and interactions between galaxies in general. Detailed studies of such individual systems, and eventually their population as a whole, can provide insight into the tidal interaction between the Magellanic Clouds, the formation and evolution of stellar populations in the Bridge, and the accretion history and dynamics of compact objects in the region.

\section*{Acknowledgements}
This work is based on data from eROSITA, the soft X-ray instrument aboard SRG, a joint Russian-German science mission supported by the Russian Space Agency (Roskosmos), in the interests of the Russian Academy of Sciences represented by its Space Research Institute (IKI), and the Deutsches Zentrum für Luft- und Raumfahrt (DLR). The SRG spacecraft was built by Lavochkin Association (NPOL) and its subcontractors, and is operated by NPOL with support from the Max Planck Institute for Extraterrestrial Physics (MPE).
The development and construction of the eROSITA X-ray instrument was led by MPE, with contributions from the Dr. Karl Remeis Observatory Bamberg \& ECAP (FAU Erlangen-Nuernberg), the University of Hamburg Observatory, the Leibniz Institute for Astrophysics Potsdam (AIP), and the Institute for Astronomy and Astrophysics of the University of Tübingen, with the support of DLR and the Max Planck Society. The Argelander Institute for Astronomy of the University of Bonn and the Ludwig Maximilians Universit{\"a}t Munich also participated in the science preparation for eROSITA. 
The eROSITA data shown here were processed using the eSASS software system developed by the German eROSITA consortium.
This work is based on observations obtained with \xmm, an ESA science mission with instruments and contributions directly funded by ESA Member States and NASA. This research has made use of data and/or software provided by the High Energy Astrophysics Science Archive Research Center (HEASARC), which is a service of the
Astrophysics Science Division at NASA/GSFC.

This work has made use of data from the Asteroid Terrestrial-impact Last Alert System (ATLAS) project. The Asteroid Terrestrial-impact Last Alert System (ATLAS) project is primarily funded to search for near earth asteroids through NASA grants NMNRAS.473.4937SN12AR55G, 80NSSC18K0284, and 80NSSC18K1575; byproducts of the NEO search include images and catalogs from the survey area. This work was partially funded by Kepler/K2 grant J1944/80NSSC19K0112 and HST GO-15889, and STFC grants ST/T000198/1 and ST/S006109/1. The ATLAS science products have been made possible through the contributions of the University of Hawaii Institute for Astronomy, the Queen’s University Belfast, the Space Telescope Science Institute, the South African Astronomical Observatory, and The Millennium Institute of Astrophysics (MAS), Chile.
This work makes use of data from the All-Sky Automated Survey for Supernovae (ASAS-SN), which is supported in part by the Gordon and Betty Moore Foundation and the Alfred P. Sloan Foundation.
The OGLE project has received funding from the Polish National Science Centre grant OPUS-28 2024/55/B/ST9/00447 awarded to AU.
The spectroscopic observation reported in this paper was obtained with the Southern African Large Telescope (SALT).

This research has made use of the VizieR catalogue access tool, CDS, Strasbourg, France (DOI:10.26093/cds/vizier) \citep{ochsenbein2000}. 

This work made extensive use of the following \texttt{python} packages: \texttt{NumPy} \citep{numpy}, \texttt{Matplotlib} \citep{matplotlib}, \texttt{SciPy} \citep{scipy}, \texttt{pandas} \citep{pandas1, pandas2}, and \texttt{Astropy}\citep{astropy2022}.

PAC would like to thank Patricia Whitelock for useful discussions on evolved symbiotic systems, and also acknowledges the Leverhulme Trust for an Emeritus Fellowship.
\section*{Data Availability}
The public X-ray data used in the work are available through the High Energy Astrophysics Science Archive Research Center (HEASARC) archive. The optical photometric data is available through the respective archives. 
Other datasets presented in the paper are available upon reasonable request.



\bibliographystyle{mnras}
\bibliography{j0431-71v4}

\newpage
\appendix
\section{Infrared colour-magnitude diagram}\label{apdx:ir-cmd}
\begin{figure}
    \centering
    \includegraphics[scale=0.5]{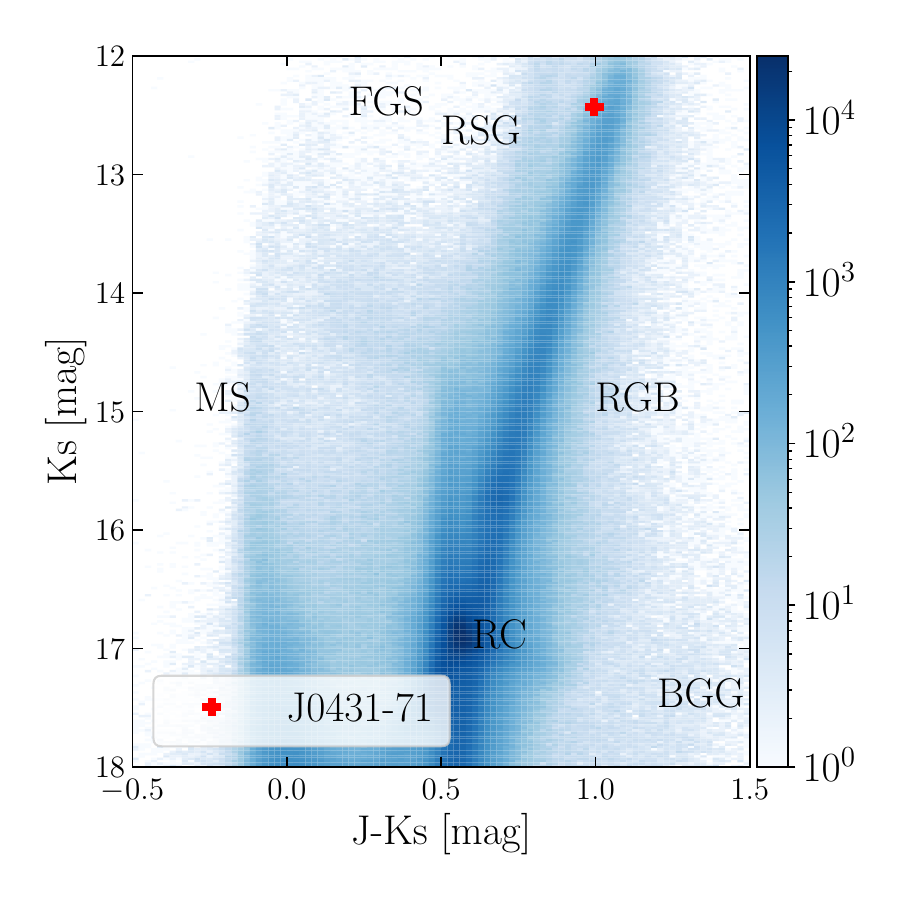}
    \caption{Colour-magnitude diagrams $Ks$ vs $J-Ks$ in the infrared band.}
    \label{fig:ircmd}
\end{figure}

The infrared colour magnitude diagram is plotted in figure \ref{fig:ircmd}. The position of J0431-71 in the diagram is consistent with the stellar companion likely being a red-giant.

\section{Periodograms of optical light-curves}\label{apdx-lombscargleothers}
\begin{figure}
\centering
    \includegraphics[scale=0.58]{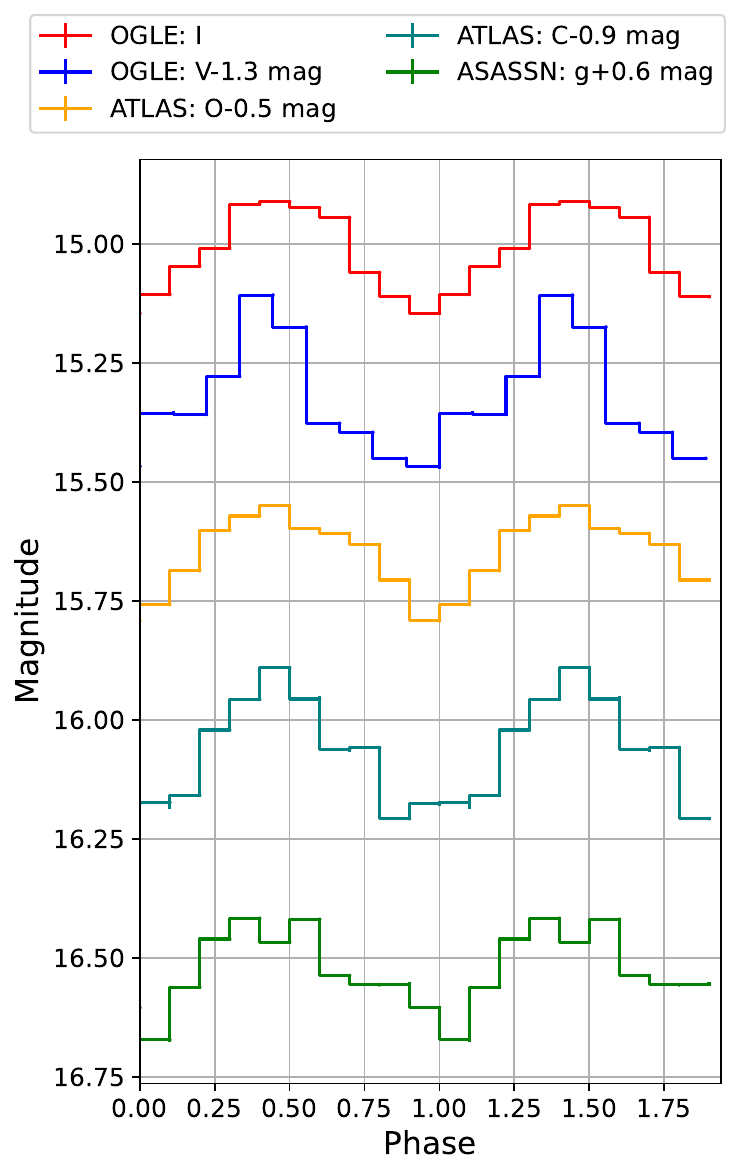}
    \caption{Phased  $I$, $V$, $o$, $c$ and $g$-band light-curves from OGLE, ATLAS and ASAS-SN, folded with the period (524 d) calculated from the OGLE-$I$ band light-curve (Table \ref{tab:lsperiodogram}).}
    \label{fig:ogleIVphase}
\end{figure}

\begin{figure*}
    \centering
    \gridline{\fig{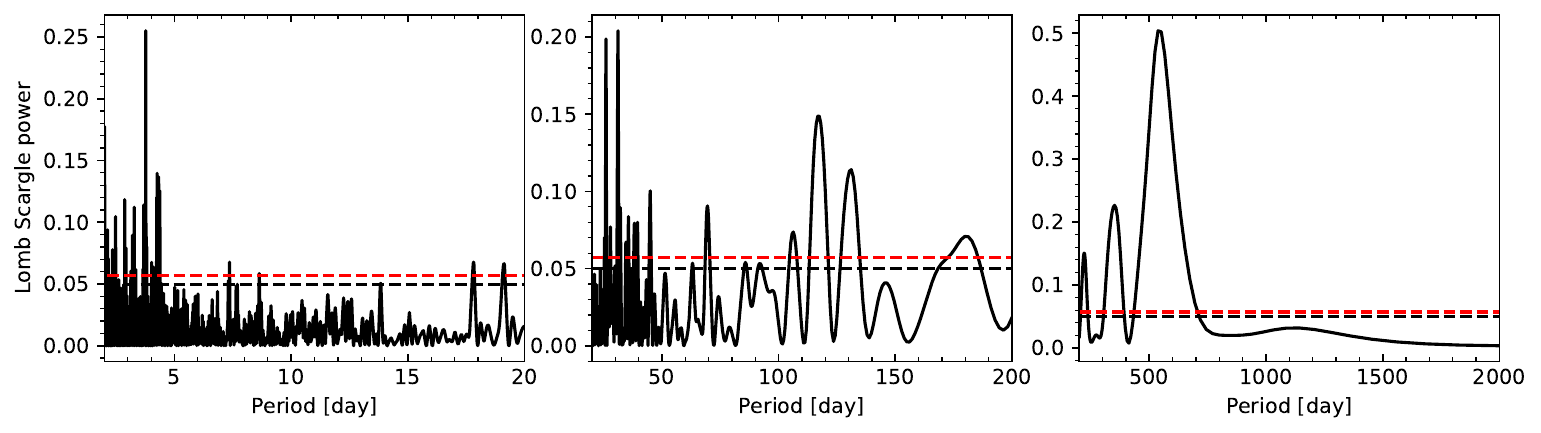}{0.92\textwidth}{(a)}}
                      \gridline{\fig{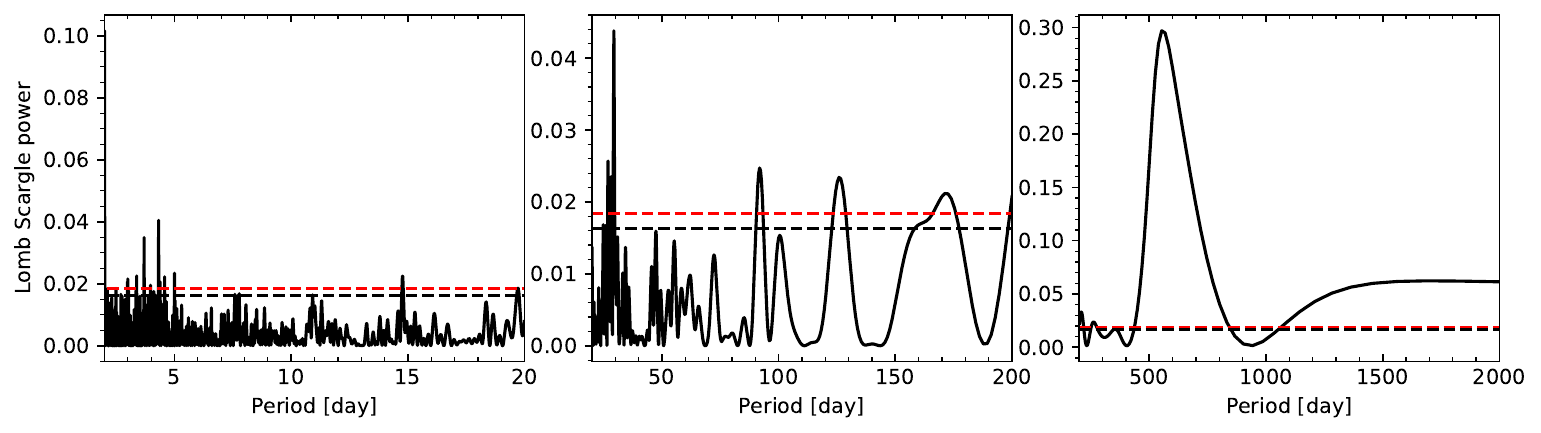}{0.92\textwidth}{(b)}}
    \gridline{\fig{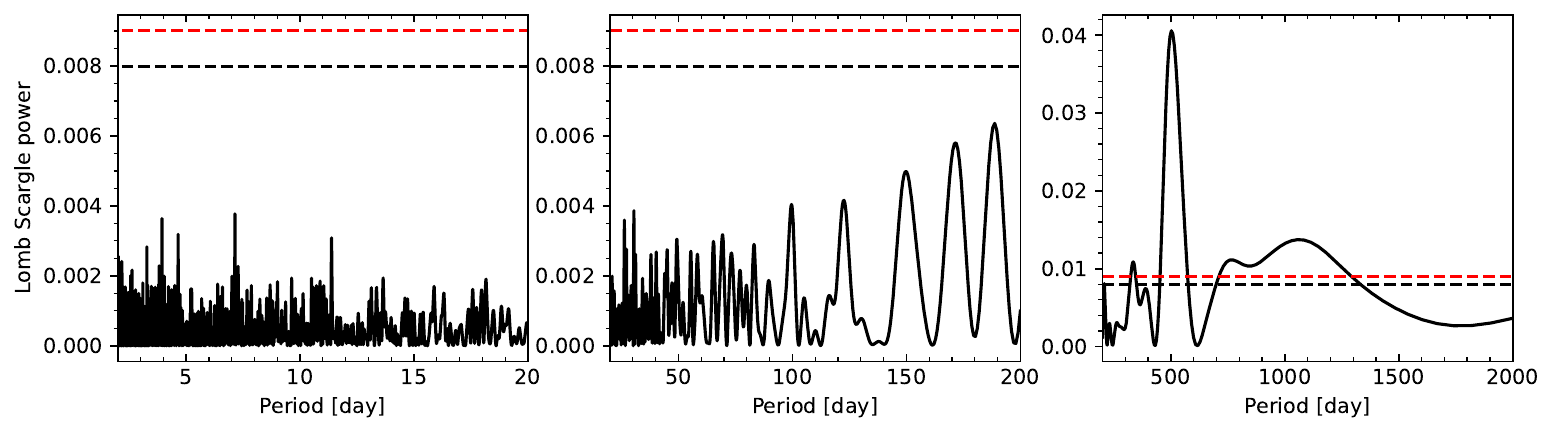}{0.92\textwidth}{(c)}}
    \caption{LS periodogram (a) ATLAS-$c$ (b) ATLAS-$o$, (c) ASAS-SN-$g$. left to right -- periodogram with period($1/f$) in 2-20 days, 20-200, and 200-2000 days interval.}
    \label{fig:LSperiodogramOthers}
\end{figure*}

We plot the OGLE phased I and V band light-curves (Figure \ref{fig:ogleIVphase}). The correlation between the V- and I- bands indicate the stellar origin of the variability in the IR-Optical bands. The broadness of the IR- peaks compared to the optical peaks might indicate that the IR-is rather emitted from an extended area. The optical light curve periodograms from ATLAS and ASAS-SN are presented in figures \ref{fig:LSperiodogramOthers}.

\section{Corner plots from BXA}\label{apdx-bxacorners}
\begin{figure*}
    \centering
    \includegraphics[scale=0.4]{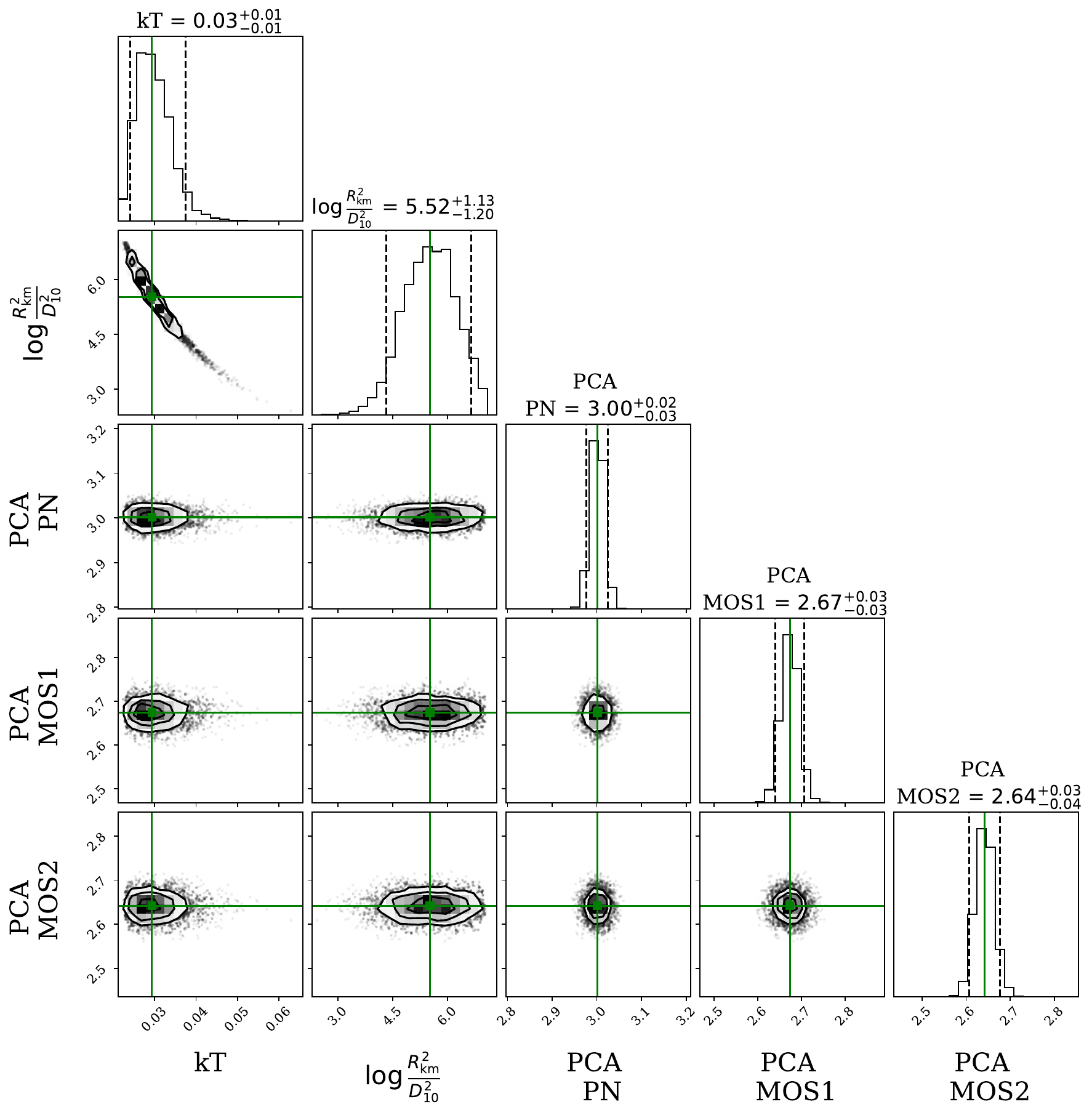}
    \caption{Corner plot of BXA-posteriors from \xmm spectral fitting.}
    \label{fig:xmmposterior}
\end{figure*}

\begin{figure*}
\centering
    \gridline{
\fig{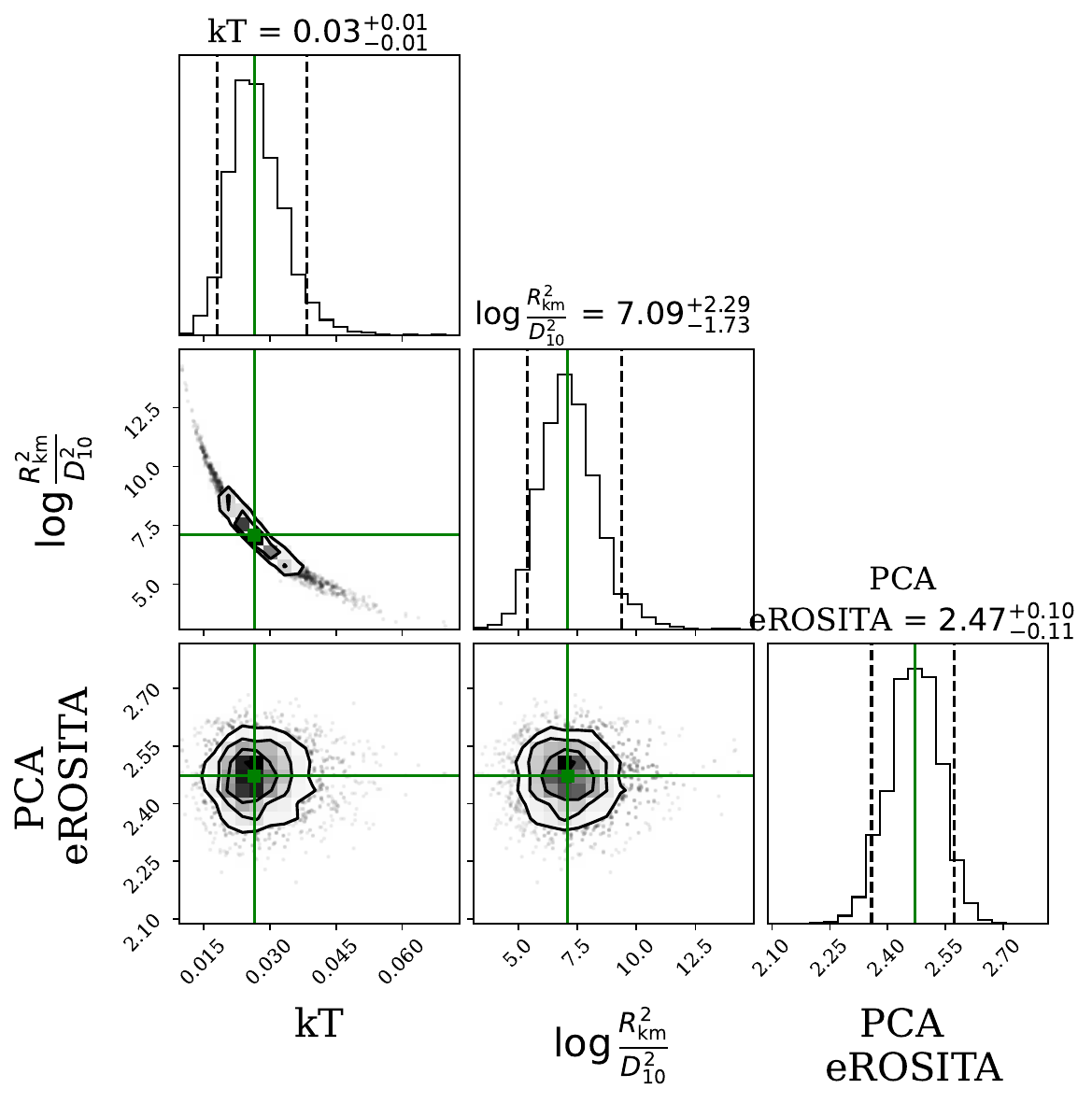}{0.5\textwidth}{(a)}    \fig{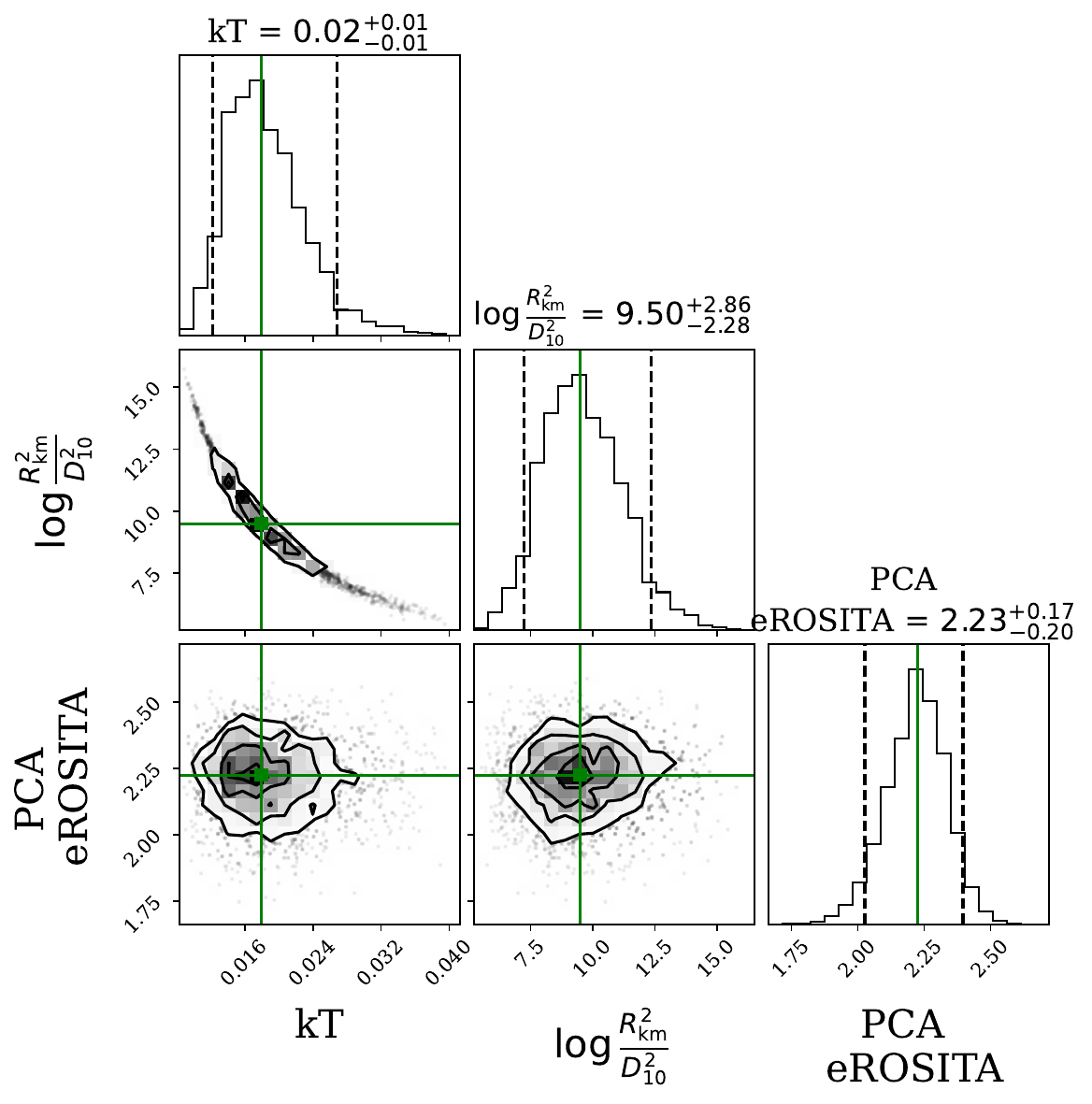}{0.5\textwidth}{(b)}
    }
    \caption{Corner plot of BXA-posteriors from (a) eRASS2 and (b) eRASS3 spectral fitting.}\label{fig:eRposterior}
\end{figure*}

The corner plots from our BXA spectral fitting (Section \ref{sec:xray-spectral-analysis}) to the \xmm and \eR data are presented in figures \ref{fig:xmmposterior} and \ref{fig:eRposterior}. The temperature $kT$ and the normalization representative of the radius of the blackbody highly degenerate and is negatively correlated. Overall, the large errors do not allow us to track changes in the temperature values across the X-ray observations.

\section{RLOF triggered accretion and the observed period}\label{sec:apdx-accretion-timescale}
The accretion flow onto the white dwarf surface is initiated by the 
circularization of material transferred via Roche lobe overflow (RLOF). Within 
each brightness variation cycle of 524\,d, the surface mass density of 
the accretion disk is modulated i.e. elevated during the bright state and 
substantially reduced during the faint state.
This accretion timescale should be consistent with the viscous timescale, $t_{\rm visc}$,  
of the disk evaluated at the circularization radius.
It is given by
\begin{equation}
    t_{\rm visc} = \frac{(R/H)^2}{\alpha}\sqrt{\frac{R^3}{GM}},
\end{equation}
where $\alpha$ is the Shakura-Sunyaev viscosity parameter 
\citep{ss73}. Adopting
$M_{WD} = 1.0\,M_{\odot}$, a donor mass in the range $1.0$--$5.0\,M_{\odot}$, 
and disk parameters appropriate for a hot, irradiated accretion disk 
($\alpha = 0.5$, $H/R = 0.5$), then $t_{\rm visc}$ evaluated at the 
circularization radius spans $430$--$634$\,d, encompassing the observed 
pulsation period of 524\,d. These values of $\alpha$ and $H/R$ are 
consistent with a slim, radiation-pressure supported disk 
\citep{abramowicz1988} expected at the inferred accretion rate of $\dot{M} \sim 
10^{-6}\,M_\odot\,{\rm yr}^{-1}$, which approaches the Eddington limit 
for a $1.0\,M_\odot$ WD.
This agreement supports the interpretation that the 524\,d brightness modulation 
reflects the timescale of the accretion flow formation and depletion, and is driven by the 
periodic variation 
in mass transfer induced RLOF by the donor pulsation.

\section{Estimation of $\beta_{\rm OX}$}\label{sec:apdx-betaox-methods}
\begin{figure}
    \centering
    \includegraphics[scale=0.60]{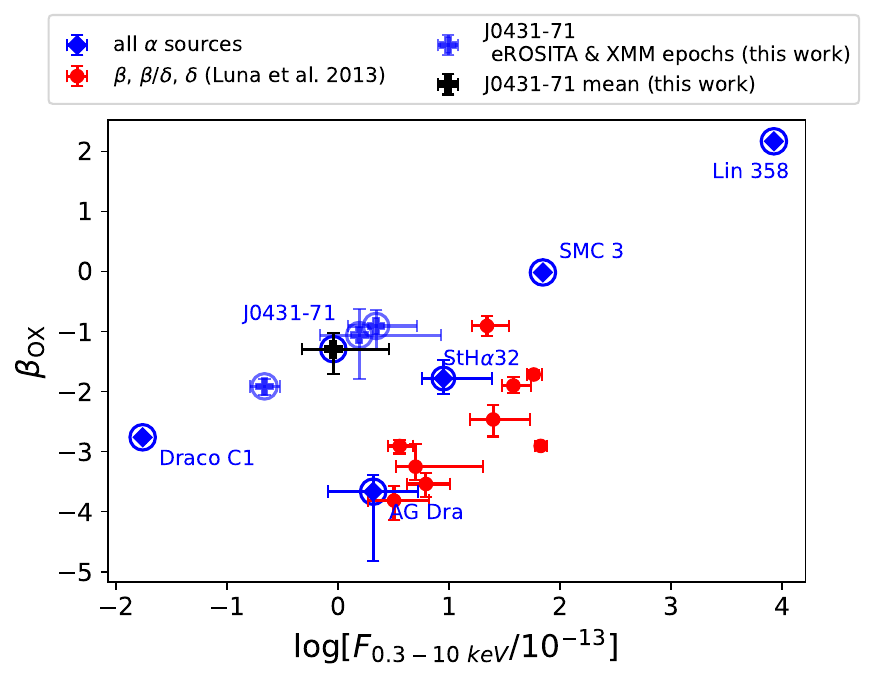}
    \caption{Plot of $\beta_{\rm OX}$ vs 0.3--10~keV flux for X-ray bright symbiotics. These include J0431-71 from three epochs (eRASS2, eRASS3, and XMM1) and their average value (this work), SMC-3 and Lin 358 with $\alpha$-type spectrum from \citet{orio2007}, and sources from Table 2 of \citet{luna2013} except Swift\,J171951.7-300206. $\beta_{\rm OX}$ is plotted for two different values of Hen 3-461's flux, which correspond to two different phenomenological fitting models \citep[Table 2 of][]{luna2013}.}
    \label{fig:fxfopt}
\end{figure}
 A quantitative empirical measure of the broadband spectral properties of accreting sources
is the optical-to-X-ray spectral slope, $\beta_{\rm OX} = \log(f_{\rm X}/f_{\rm opt})$, originally 
introduced by \cite{maccacaro1988}. In this work, we adopt a modified definition of $\beta_{\rm OX}$ by 
replacing the V-band magnitude with the average of the \textit{Gaia} BP and RP magnitudes. The modified expression is given by:

\begin{equation}\label{eq:betaOX}
\beta_{\rm OX} = \log f_{\rm X,0.3-10~keV} + \left[ \frac{BP+RP}{5} \right] + 5.37
\end{equation}

For comparison with J0431-71, we compiled symbiotic SSS from the literature, 
including nine sources from \cite{luna2013} with reported unabsorbed 0.3--10~keV fluxes, Lin~358 and SMC~3 
from \cite{orio2007}, AG Dra from \citep{skopal2009}, and Draco~C1 from \cite{saeedi2018}. We obtain the $BP$ and $RP$ magnitudes from GAIA eDR3 \citep{gaia2021}.

We find that the symbiotic sources form two distinct branches in the $\beta_{\rm OX}$ distribution. The lower, steeper branch contains all $\beta$, $\beta/\delta$, and $\delta$ sources, along with a few $\alpha$ systems like AG-Dra, while the upper, flatter branch is dominated by $\alpha$-type sources (Figure \ref{fig:fxfopt}). A more detailed study using a consistent optical band and X-ray energy range is required to investigate this trend further. Nevertheless, the overall distribution suggests that these systems exhibit a harder-when-brighter behavior.

\bsp	
\label{lastpage}
\end{document}